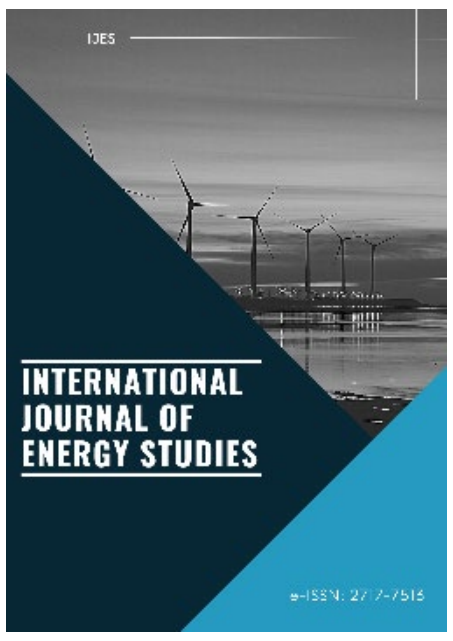

# Assessment and prioritization of renewable energy alternatives to achieve sustainable development goals in Türkiye: Based on fuzzy AHP approach[1]

**Emre Akusta[a*], Raif Cergibozan[b]**

*[a]Department of Economics, Kirklareli University, 39100, Kirklareli, Türkiye, ORCID: 0000-0002-6147-5443*
*[b]Department of Economics, Kirklareli University, 39100, Kirklareli, Türkiye, ORCID: 0000-0001-7557-5309*
*(*Corresponding Author:emre.akusta@klu.edu.tr )*

**Highlights**

- This study prioritizes renewable energy sources to achieve sustainable development in Türkiye using the fuzzy AHP method.
- We analyze five main criteria, thirty sub-criteria and five renewable energy sources.
- We also analyzed which renewable energy source is the best option for each main and sub-criteria.
- The results show that the most important main criteria for renewable energy investments in Türkiye are economic, political, technical, environmental and social criteria, respectively.
- The most suitable renewable energy sources for Türkiye are solar, wind, hydroelectric, biomass and geothermal, respectively.

## ABSTRACT

The aim of this study is to prioritize renewable energy sources to achieve sustainable development in Türkiye by using fuzzy AHP method. In our study, we used 30 criteria that affect the investment in renewable energy sources. We also calculated the weights of these criteria in investment decisions. In addition, we analyzed the advantageous renewable energy sources according to each criterion. Thus, it was determined which renewable energy source is advantageous according to which criteria. The results show that the most important main criteria for renewable energy investments in Türkiye are economic, political, technical, environmental and social criteria, respectively. The most appropriate renewable energy sources according to economic, political, technical and social criteria are solar, wind, hydroelectric, biomass and geothermal respectively.

*[1] This study is derived from Emre Akusta's PhD thesis prepared under the supervision of Prof. Raif Cergibozan.*

## 1. INTRODUCTION

Currently, it is widely accepted that the growth policies pursued since the industrial revolution are no longer sustainable. Because the pressure on the environment caused by production based on fossil resources has increased to levels that cannot be ignored and endanger future generations. In addition, the economic growth efforts of developing countries further increase the global pressure on natural resources [1]. Energy use cannot be abandoned despite increasing environmental pressures. Because energy is very important not only for industry but also for daily life. The level of civilization advances with energy use. Therefore, the total amount of energy demanded is increasing every day. Countries also need energy to maintain their current level of welfare. Countries do not want to compromise the level of prosperity they have reached. Therefore, the amount of energy consumed is not decreasing and it is estimated that it will not decrease in the future. Indeed, projections for 2050 indicate that fossil fuels will continue to be used to meet most of the world's energy demand [2]. As a result, natural resources such as soil, air and water are becoming more and more polluted and threatening life [3]. Thus, the perspective that development can only be measured by economic growth has been questioned and the concept of sustainable development has emerged. Because this decision also has economic, social and environmental impacts. Hence, only giving importance to one of these impacts in decision-making has unsustainable consequences. In other words, sustainable development has three dimensions: economic, social and environmental [4]. Environmental sustainability focuses on the quality of the environment and natural resources necessary to meet people's needs, maintain quality of life and sustain economic activities. Social sustainability focuses on issues such as human rights, social equality and social justice. Economic sustainability focuses on the protection of natural, social and human capital, economic growth and economic stability. To achieve sustainable development, there must be harmony between these dimensions and successfully balance them. One dimension of sustainable development should not undermine the other [5].

Sustainable development is an important goal for both developed and developing countries. It is especially important for developing countries. Developing countries face challenging problems such as tackling environmental problems, reducing poverty and sustaining economic growth. On the other hand, climate change, global warming and the destruction of the ozone layer are among the prominent problems in sustainability [4], [6]. For a sustainable environment, rational use of renewable energy resources and limited use of fossil energy resources are required [7]. Since the economy and the environment are closely related, it is impossible to consider the environment separately from the economy. Therefore, effective use of natural resources and the environment is very important for sustainable economic growth and sustainable development. Because there is a very close relationship between environmental pollution, natural resources and economy. Increased economic activity leads to environmental problems, and increased environmental problems damage the economy and infrastructure. Thus, while economic activities cause environmental pollution, environmental pollution leads to increased economic and social costs.

The increasing importance of sustainable development has also brought the measurement methods of this concept to the agenda. Early studies in the literature measured sustainable development with Gross Domestic Product (GDP) per capita. However, over time, it was realized that this single indicator was insufficient to represent all dimensions of development. In order to overcome this deficiency, the Human Development Index was created by adding indicators that increase physical and social welfare. In later studies, indicators such as education, health, resource consumption and environmental degradation were added to the Human Development Index to measure sustainable development more comprehensively [8]. More recently, energy and environmental indicators such as per capita electricity consumption, energy intensity and carbon emissions have been included in the Human Development Index calculations [9]. Because energy is at the center of economic, social and environmental dimensions of sustainable development [10]. At this point, the importance of renewable energy sources emerges. Unlike fossil fuels, renewable energy sources are inexhaustible and their destructive impact on the environment is very limited. In addition, renewable energy resources can contribute to sustainable energy and energy security [11]. Investing in renewable energy resources in Türkiye is expected to increase both energy supply security and support environmental sustainability. Therefore, in our study, we analyzed the most suitable renewable energy source to be invested in Türkiye for sustainable development.

This study can contribute to the literature in at least four ways: (1) To the best of our knowledge, the number of studies prioritizing renewable energy resources for sustainable development in Türkiye is quite limited. This study aims to improve the literature on this field. (2) A wider set of criteria was used compared to previous studies. This allows for a more comprehensive approach to the evaluation of renewable energy sources. The use of 30 criteria under 5 dimensions expands the scope of the analysis in the literature and provides a more detailed perspective. This set of criteria, which is based on expert opinions, allows for a more in-depth evaluation of decision-making processes in the field of renewable energy. (3) By using the Fuzzy AHP method, uncertainties and subjective judgments are addressed more sensitively compared to the classic AHP method in the literature. This method makes a methodological contribution to the literature with the use of fuzzy logic in the field of sustainable development. (4) The sensitivity analysis conducted in the study tested the validity of the results and evaluated the robustness of the findings. This analysis provides another methodological contribution to the literature by revealing the reliability of the model and the impact of variables in the decision-making process.

The rest of the study is organized as follows. Section 2 presents the literature review, Section 3 the data and methodology, Section 4 the results and discussion, Section 5 the policy implications and Section 6 the conclusion.

## 2. LITERATURE REVIEW

The environmental problems caused by energy use resulting from the increase in production and consumption have made it clear that the current situation is unsustainable. Therefore, the necessity of a sustainable world order has come to the agenda. It is thought that renewable energy resources can be a solution to ensure sustainability. For this reason, the number of collaborations and academic studies on sustainable development and renewable energy is increasing with each passing day.

The impacts of renewable energy resources on sustainable development have been addressed from various angles in the literature. Batı [11] examined the contribution of renewable energy use to sustainable development in Türkiye. In particular, it is emphasized that the state should invest to ensure energy security. The study shows that the inexhaustibility and abundance of renewable energy resources are among the reasons why these resources are preferred. Similarly, Fotis and Polemis [12] investigated the relationship between sustainable development and renewable energy in European Union countries. In the study, it is stated that energy intensity increases pollution, but energy saving reduces environmental degradation. It is indicated that new technologies and renewable energy sources will support sustainable development. Güney [13] compared the impacts of renewable and fossil energy sources on sustainable development in developed and developing countries. The results revealed that renewable energy sources have a positive impact on sustainable development in both country groups. It is emphasized that the level of sustainable development increases with the increase in the amount of renewable energy compared to fossil fuels. This study reinforced the importance of renewable energy sources, especially in achieving the 2030 Sustainable Development Goals. Dinçer and Karakuş [14] analyzed the impacts of renewable energy on sustainable economic development for BRICS and MINT countries. They found that renewable energy use is one of the most important components of economic development in Brazil and China. It was concluded that more attention should be paid to renewable energy investments among these countries. Öymen [15] interviewed experts working in the energy sector. The results revealed that domestic renewable energy sources increase sustainability. However, the importance of government support for the development of renewable energy was also emphasized. Finally, Tiba and Belaid [16], in their study in African countries, stated that renewable energy investments play an important role in achieving the UN's 2030 Sustainable Development Goals. They emphasized that renewable energy makes great contributions in terms of providing easier access to energy, reducing environmental pollution and supporting sustainable development. The study also clearly demonstrates the economic, social and environmental benefits of renewable energy.

The literature emphasizes the important role of renewable energy sources on sustainable development. It also highlights the importance of developing policies and incentives appropriate to the dynamics of each region and country. In light of this literature emphasizing the importance of renewable energy sources, it becomes clear that energy investments should be made by considering not only environmental but also political, economic and social balances. At this point, decision-making processes for energy investments are quite complex.

Energy investments involve many choice problems. These decisions require a balance in many areas including political, environmental, economic and social. Therefore, mathematical methods have started to be used in the solution of choice problems in the energy field. In this section, we review studies that analyze energy resources using multi-criteria decision making techniques. Among these studies, Kahraman et al. [17] ranked the most suitable renewable energy sources for Türkiye as wind, solar, biomass, geothermal and hydroelectric. Similarly, Kahraman and Kaya [18] also emphasize wind and solar energy. Atmaca and Basar [19] concluded that the best alternative energy source for Türkiye is nuclear energy, followed by natural gas, geothermal, wind, hydroelectricity and coal. Demirtas [20] aimed to determine the most appropriate renewable energy technology for renewable energy planning using the AHP method. The results of the study showed that the priority ranking of renewable energy sources is wind, biomass, geothermal, solar and hydroelectric energy. However, some studies (see for example [21], [22], [23]) have identified hydropower as the most suitable energy source for Türkiye.

A group of studies examined alternative energy sources in Türkiye from a broader perspective. Celikbilek and Tuysuz [24] integrated DEMATEL, AHP and VIKOR methods in their study. The results of the study show that energy sources are solar, wind, hydroelectric, biomass and geothermal. Sagir and Doganalp [25] evaluated renewable energy resources by considering criteria such as reliability, cost, risk and contribution to the national economy. The results revealed that renewable energy sources are more suitable than nuclear energy sources and fossil energy sources. Balin and Baracli [26] evaluated wind, solar, biomass, geothermal, hydroelectric and hydrogen energy sources using fuzzy AHP method. As a result, they determined wind energy as the most suitable option. Buyukozkan and Guleryuz [27] also analyzed the most suitable renewable energy sources in Türkiye. In the study, DEMATEL, ANP and TOPSIS methods were used and 20 sub-criteria under 5 main criteria were considered. Unlike many other studies, they found that the best renewable energy sources for Türkiye are geothermal and biogas. These results can be associated with the legal difficulties in wind power plants and the poor environmental performance of hydroelectric power plants in those years. Colak and Kaya [28] used fuzzy AHP and fuzzy TOPSIS methods and found that wind is the most suitable renewable energy source for Türkiye. Ozcan et al. [29] determined that wind is the most suitable renewable energy source for Türkiye after hydroelectricity. Ozkale et al. [30], who used the PROMETHEE method for the same purpose, concluded that hydroelectric energy is the most suitable source. Boran [31] used the fuzzy VIKOR method. The results showed that the ranking of renewable energy sources is wind, hydroelectricity, solar energy.

Renewable energy sources have also been examined in terms of economic and environmental sustainability. Büyüközkan et al. [32] and Karaca and Ulutaş [33] investigated the most suitable energy sources for Türkiye to increase economic and environmental sustainability. The results show hydropower as the most suitable energy source. Engin at el. [34], who evaluated Türkiye's energy alternatives, reported that solar energy is prioritized. Toklu and Taskin [35] determined wind energy as the most suitable energy source for Türkiye using the same methods. Karakas and Yildiran [36] evaluated Türkiye's renewable energy alternatives using the Modified Fuzzy AHP method. The results demonstrate that solar energy is the most suitable option. Morever, Derse and Yontar [37] analyzed Türkiye's energy resources by combining SWARA and integrated TOPSIS methods. The results show that biomass energy is the most suitable source. Evaluating Türkiye's electricity generation technologies, Yilan et al. [38] identified hydroelectric power plants with dams as the most suitable option. Solangi et al. [39] determined wind as the most suitable energy source for Türkiye by using Delphi, Fuzzy AHP and Fuzzy WASPAS methods. Similarly, Deveci et al. [40] found the same result using the fuzzy CODAS method. Karatop et al. [41] evaluated five renewable energy alternatives using 13 criteria in their study. The study concluded that Türkiye should focus on hydroelectricity and wind energy in its renewable energy investments. Finally, Bilgili et al. [42] analyzed the best renewable energy options for Türkiye using the IF-TOPSIS method. The analysis results indicate that solar energy is the best renewable energy source for Türkiye's sustainable growth. In addition, the most important criterion affecting renewable energy investments in Türkiye is the investment cost.

We reviewed studies that used multi-criteria decision-making methods to select the most appropriate energy source to meet Türkiye's energy needs. Most of these studies selected renewable energy sources, while some included fossil energy sources. This difference varies according to the priorities of the researchers and the current situation in the energy sector. It was also observed that most of the studies used 4 main criteria: economic, technical, social and environmental factors. The number of sub-criteria varies between 8 and 35. The number of sub-criteria varies depending on the scope and objectives of the study. VIKOR, AHP, Fuzzy AHP, TOPSIS and MACBETH methods are the most used methods in these studies. Wind, solar and hydroelectricity are the most dominant energy sources in most of the studies. It was also found that wind and solar are generally ranked first. In addition, international studies on renewable energy source selection are shown in Table 1.

**Table 1.** International literature on the selection of renewable energy source

| Authors | MCDM method | Country | Energy sources | Number of criteria | Conclusion |
|---|---|---|---|---|---|
| San Cristoball [43] | VIKOR | Spain | Hydropower, Solar, Biomass, Wind. | 7 | Biomass |
| Yi et al. [44] | AHP | North Korea | Wind, Geothermal, Hydropower, Solar, Biomass. | 10 | Wind |
| Sadeghi et al. [45] | AHS Fuzzy TOPSIS | Iran | Solar, Wind, Hydropower, Geothermal. | 13 | Solar |
| Mourmouris and Patolias [46] | REGIME | Greece | Solar, Wind, Hydropower, Geothermal. | 16 | Wind |
| Stojanović [47] | AHP | - | Solar, Biomass, Geothermal, Hydropower, Wind. | 12 | Wind |
| Ahmad and Tahar [48] | AHP | Malaysia | Hydropower, Solar, Wind, Biomass. | 12 | Solar |
| Troldborg et al. [49] | PROMETHEE | Scotland | Solar, Wind, Hydropower, Geothermal. | 9 | Solar |
| Tasri and Susilawati [50] | Fuzzy AHP | Indonesia | Biomass, Wind, Geothermal, Hydropower, Solar. | 16 | Hydropower |
| Al Garni et al. [51] | AHS | Saudi Arabia | Solar, Wind, Biomass, Geothermal. | 14 | Solar |
| Afsordegan et al. [52] | Fuzzy AHP, TOPSIS | - | Nuclear, Hydropower, Solar, Wind, Geothermal, Biomass. | 9 | Wind |
| Robles et al. [53] | AHP | Colombia | Hydropower, Solar, Wind, Biomass. | 20 | Solar |
| Ishfaq et al. [54] | TOPSIS, AHP | Pakistan | Hydropower, Solar, Wind, Biomass. | 6 | Hydropower |
| Yuan et al. [55] | HFLTS | China | Solar, Biomass, Hydropower, Wind. | 10 | Biomass |
| Rani et al. [56] | VIKOR | India | Hydropower, Solar, Wind, Biomass. | 13 | Wind |
| Solangi et al. [57] | Delphi, AHP, Fuzzy TOPSIS | Pakistan | Wind, Geothermal, Hydropower, Solar, Biomass. | 20 | Wind |
| Rani et al. [58] | Fuzzy TOPSIS | India | Nuclear, Hydropower, Solar, Wind, Geothermal, Biomass. | 9 | Wind |
| Niu et al. [59] | Fuzzy ELECTRE II | China | Solar, Biomass, Geothermal, Hydropower, Wind. | 13 | Hydropower |
| Li et al. [60] | Fuzzy VIKOR, Fuzzy TOPSIS | - | Solar, Biomass, Geothermal, Hydropower, Wind. | 8 | Wind |
| Chen et al. [61] | PROMETHEE II | China | Solar, Biomass, Hydropower, Wind. | 12 | Solar |
| Wang et al. [62] | Fuzzy AHP | Pakistan | Solar, Wind, Biomass. | 17 | Wind |
| Mohammed et al. [63] | AHP | Iraq | Solar, Geothermal, Hydropower. | 7 | Solar |
| Abdul et al. [64] | AHP, VIKOR | Pakistan | Solar, Biomass, Hydropower, Wind. | 16 | Hydropower |
| Assadi et al. [65] | Fuzzy Delphi | Iran | Biomass, Wind, Solar, Geothermal, Hydropower, Marine, Hydrogen. | 12 | Solar |
| Goswami et al. [66] | MEREC, PIV | India | Solar, Biomass, Geothermal, Hydropower, Wind. | 6 | Hydropower |
| Li et al. [67] | DEMATEL, PROMETHEE | China | Geothermal, Nuclear, Solar, Rüzgar, Hydropower . | 5 | Nuclear |

Source: Authors' construction.

Table 1 presents studies conducted in different countries on renewable energy source selection. The results of the previous literature are in line with the studies conducted in Türkiye. The most emphasized energy sources in most of the studies are wind, solar and hydroelectricity. It is seen that wind and solar are generally ranked first among renewable energy sources.

## 3. DATA AND METHODOLOGY

### 3.1. Selection of Indicators

The AHP hierarchy has the objective at the top, the main criteria below it, and sub-criteria under each of the main criteria. Alternatives are at the bottom of the hierarchy. In our study, the objective was determined first. The objective of our study is to determine the most suitable renewable energy source that should be invested in Türkiye for sustainable development. The alternatives used in our study are renewable energy sources such as Wind, Solar, Geothermal, Biomass and Hydroelectric. In the next stage, it is necessary to determine the main and sub-criteria to be used in the AHP method. The criteria were selected from previous literature. The criteria to be used in the study are shown in Table 2. The expanded version of Table 2 with references is provided in Appendix A.

Table 2 shows the main criteria and sub-criteria used in the study collectively. It is widely accepted in the literature that different criteria are emphasized among countries in the selection of renewable energy sources. Since the socio-economic structure, energy policies, environmental objectives and resource potential of each country are different, the selection of renewable energy sources also varies depending on these factors. The criteria used in this study are compiled from studies specific to Türkiye. In addition, each of the criteria used in the study is associated with the Sustainable Development Goals (SDGs). These SDGs are shown in the last column of the table. Explanations of the criteria used in the study are also included in the table. After determining the criteria to be used in the study, the Fuzzy AHP method is explained in the following section.

**Table 2.** Criteria for renewable energy selection

| Criteria | Sub-criteria | | Description | SDG |
|---|---|---|---|---|
| $C_1$. Technical | $C_{11}$ | Efficiency | Efficiency of primary energy conversion to electricity. | 7-8-9-12 |
| | $C_{12}$ | Technical risk | Risks and hazards in the energy production process. | 7-11 |
| | $C_{13}$ | Capacity factor | Maximum power capacity that the power plant can generate and accommodate. | 7-9 |
| | $C_{14}$ | Technology maturity | How widespread the technology used is at regional, national and international level. | 7-8 |
| | $C_{15}$ | Resource potential | Availability of resources used to produce energy. | 7 |
| | $C_{16}$ | Implementation speed | The period from project phase to operation. | 7 |
| | $C_{17}$ | Operational life | The period during which the power plant can operate efficiently. | 7-9 |
| | $C_{18}$ | Ease of implementation | Simplicity of the power plant and its technology. | 7 |
| | $C_{19}$ | Ease of access to resources | Ease of access to the resource to be used in energy production. | 7 |
| $C_2$. Economic | $C_{21}$ | Investment cost | Initial investment cost of the power plant. | 7 |
| | $C_{22}$ | Operation and maintenance costs | Variable costs and maintenance and repair costs of the power plant. | 7 |
| | $C_{23}$ | Energy generation cost | Cost per unit of electricity generated from power plants. | 7 |
| | $C_{24}$ | Market development | Current demand and future demand potential of the power plant. | 8-17 |
| | $C_{25}$ | Contribution to national economy | Contribution of the power plant to the national economy. | 8 |
| | $C_{26}$ | Contribution to local economy | Contribution of the power plant to the local economy. | 8-11 |
| | $C_{27}$ | Continuity of energy generation | The length of time that energy production can be sustained. | 7-9 |
| | $C_{28}$ | Payback period | Payback period for initial investments. | 7-9 |
| $C_3$. Political | $C_{31}$ | Foreign dependency | Contribution to reducing foreign dependence in terms of both energy resources and implementation technologies. | 7-17 |
| | $C_{32}$ | Compliance with national agenda | The projects to be invested in are compatible with both political policies and legal procedures. | 12-13 |
| | $C_{33}$ | Compatibility with national energy policy | Compatibility of power plants with national energy policy. | 9-12-13 |
| | $C_{34}$ | Incentive mechanisms | Financial support and incentive mechanism for energy investments. | 9-12-13 |
| $C_4$. Social | $C_{41}$ | Social acceptability | Community willingness to accept a power plant in their area. | 11-12 |
| | $C_{42}$ | Job creation | Number of local jobs created at the power plant. | 1-8-10 |
| $C_5$. Environmental | $C_{51}$ | Greenhouse gas emissions | Potential of power plants to reduce carbon dioxide emissions. | 7-11-13-15 |
| | $C_{52}$ | Climate change risk | Potential of power plants to mitigate climate change. | 7-11-13-15 |
| | $C_{53}$ | Land requirement | Land requirement for the physical installation of the power plant. | 2-3 |
| | $C_{54}$ | Waste | Waste from the power plant and the need for disposal. | 3-6-11 |
| | $C_{55}$ | Ecological risk | Impact of the power plant on the environment, agricultural land and water. | 2-3-6-11 |
| | $C_{56}$ | Noise | Noise and vibration from power plant operation. | 11-12 |
| | $C_{57}$ | Continuity and predictability of resources | Sustainability and predictability of the energy sources to be used in the power plant. | 6-9-11-12 |

Source: Authors' construction.

### 3.2. Principle of Fuzzy AHP

The AHP Method is a multi-criteria decision-making method developed by Thomas L. Saaty in the 1970s. The AHP method, which is very popular among multi-criteria decision making methods, is based on pairwise comparison of criteria. The comparison of these criteria is based on expert opinion [68], [69], [70]. The AHP method is used in selection problems with one or more decision makers, many criteria and many alternatives. The general logic of AHP is to classify alternatives by pairwise comparison based on a certain criterion [71].

While constructing the hierarchy in the AHP method, there is the goal at the top, the main criteria to be used in comparisons underneath, sub-criteria under each of the main criteria and alternatives at the bottom of the hierarchy. The hierarchical structure of the AHP method used in this study is shown in Figure 1.

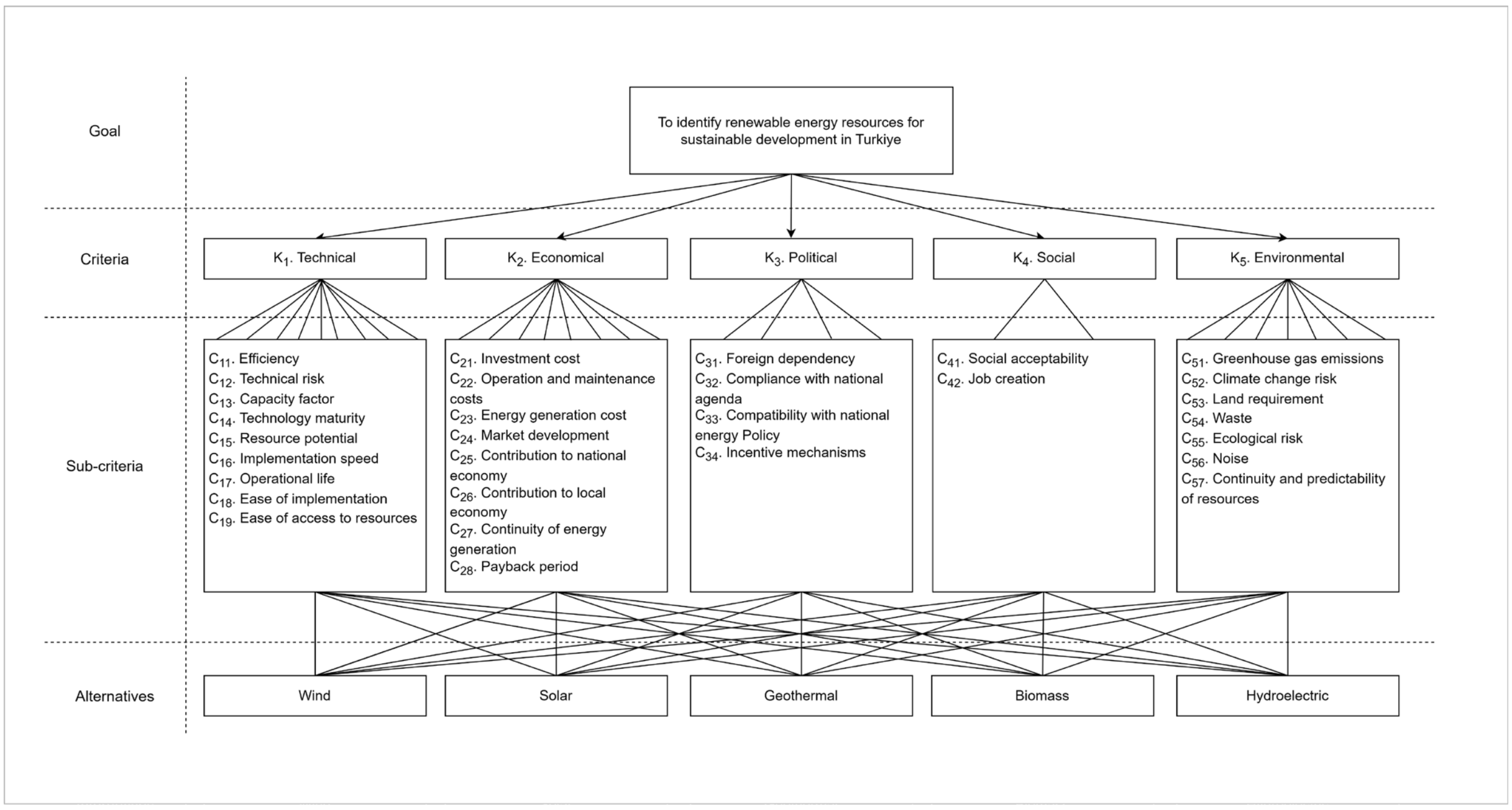

**Figure 1.** The proposed fuzzy AHP model

In the AHP method, evaluations are based on expert opinions. Therefore, it may include subjective judgments. Fuzzy AHP was developed to reduce subjectivity [72]. Fuzzy AHP is also used in this study. The decision maker ranks the criteria through pairwise comparison to determine the relative superiority of the criteria at each level. The importance scale (Table 3) proposed by Saaty [68] is used to compare the criteria and determine their importance. Decision maker makes the comparison by selecting the statement from the table that will represent his/her opinion in pairwise comparisons. In numerical calculations, numerical values selected from the table are used. Intermediate values can also be used if the decision maker is undecided about the main values in the table. In the next stage, the normalized matrix is obtained by dividing each column by the sum of the values in the relevant column.

**Table 3.** Linguistic terms and the triangular fuzzy numbers

| Precise score | Linguistic scale | Triangular fuzzy scale | Triangular fuzzy reciprocal scale |
|---|---|---|---|
| **1** | Equally important | 1,1,1 | 1,1,1 |
| **2** | Intermediate values | 1,2,3 | 1/3, 1/2, 1 |
| **3** | Weakly important | 2,3,4 | 1/4, 1/3, 1/2 |
| **4** | Intermediate values | 3,4,5 | 1/5, 1/4, 1/3 |
| **5** | Essentially important | 4,5,6 | 1/6, 1/5, 1/4 |
| **6** | Intermediate values | 5,6,7 | 1/7, 1/6, 1/5 |
| **7** | Very strongly important | 6,7,8 | 1/8, 1/7, 1/6 |
| **8** | Intermediate values | 7,8,9 | 1/9, 1/8, 1/7 |
| **9** | Absolutely important | 8,9,9 | 1/9, 1/9, 1/8 |

Source: Adapted from Saaty [68].

In the final stage, a consistency ratio (CR) is calculated for each matrix to assess the consistency of the experts' judgment. It is difficult for all pairwise comparisons or comparison matrices used in the AHP method to be one hundred percent consistent. Ultimately, these importance judgments are based on human judgments and may contain a tolerable amount of inconsistency. The Consistency Ratio (CR) was developed to determine that the inconsistency is explainable and reasonable [73].

CR is calculated by CI/RI. The first step in calculating the consistency ratio is to obtain the Nmax matrix. The Nmax matrix is calculated by multiplying the A matrix by the W matrix. The Nmax value is obtained by summing the elements of the Nmax matrix. The Consistency Indicator (CI) is obtained by processing the Nmax value as in CI=(Nmax-N)/(n-1) [74]. The last value needed to calculate the consistency ratio is the Randomness Indicator (RI). Saaty and Tran [75] prepared RI indicators for pairwise comparisons with criteria up to 15. These ratios are shown in Table 4.

**Table 4.** Values of random index (RI)

| Order | 1 | 2 | 3 | 4 | 5 | 6 | 7 | 8 | 9 | 10 | 11 | 12 | 13 | 14 | 15 |
|---|---|---|---|---|---|---|---|---|---|---|---|---|---|---|---|
| RI | 0 | 0 | 0,52 | 0,89 | 1,11 | 1,25 | 1,35 | 1,40 | 1,45 | 1,49 | 1,52 | 1,54 | 2,56 | 1,58 | 1,59 |

Source: Adapted from Saaty and Tran [75].

When the CR value is CR<1, the inconsistency is at an acceptable level. However, when CR>1, the inconsistency is above the acceptable level. Thus, decision makers need to revisit the decision matrices and eliminate the inconsistency. The stages of application of the fuzzy AHP method described above should be applied for each main and sub-criteria. The weights of the matrices whose consistency ratio is at an acceptable level are calculated. The result distribution at the decision stage is found by gathering the weights on the path that leads the alternatives in the decision hierarchy to the decision.

Fuzzy Analytic Hierarchy Process (Fuzzy AHP) was developed by combining fuzzy logic with the classical AHP method. Thus, it deals with uncertainty and subjective evaluations of decision makers in a more flexible way. In classical AHP, decision makers make precise comparisons between criteria with clear ratios. However, Fuzzy AHP was developed for situations where decision makers have difficulty in making such precise evaluations. In this way, uncertainties and inconsistencies in human judgment are more effectively managed [69], [75].

Fuzzy AHP differs from other multi-criteria decision making (MCDM) methods in that it integrates uncertainty into the model. For example, TOPSIS ranks alternatives based on their distance from the ideal solution. The closer the alternative is to the ideal solution, the better it is considered. However, TOPSIS does not consider the subjective uncertainty of the decision maker and is based on precise evaluations. Another method, VIKOR, aims to reach a compromise solution. VIKOR evaluates the differences between certain criteria when ranking alternatives. However, this method does not directly address uncertainty. Methods such as ELECTRE are based on the elimination of alternatives. ELECTRE makes comparisons between criteria to identify strong alternatives and eliminate weak ones. However, this method does not take into account the uncertainty in subjective evaluations and focuses on more precise results. PROMETHEE ranks alternatives based on preference functions determined by the decision maker. In this method, the degree of superiority of alternatives over each other is calculated. PROMETHEE, like other MCDMs, focuses on the decision maker's precise preferences. Fuzzy AHP, in contrast, allows subjective judgments to be integrated more reliably into the model by allowing each alternative to be compared in an uncertain framework [31], [32], [38], [71].

As a result, the most important difference of Fuzzy AHP compared to other MCDMs is that it can better manage the uncertainties and fuzziness in decision makers' judgments. Methods such as TOPSIS, VIKOR, ELECTRE and PROMETHEE aim to achieve specific and sharp results. Whereas Fuzzy AHP deals with uncertainty better and allows for a more flexible evaluation of alternatives. In this respect, Fuzzy AHP stands out as a more useful method when uncertainty plays an important role in complex decision processes. Therefore, Fuzzy AHP method is employed in this study.

## 4. RESULTS AND DISCUSSION

### 4.1. Determining the Weight of Criteria

In order to determine the ranking of renewable energy sources, the weights of the main criteria and sub-criteria need to be determined. We consulted 13 experts in our study and their profiles are shown in Appendix B. The matrices (and CR values) created for the main criteria and sub-criteria used in the study are presented in Appendix C. The weight coefficients and importance ranking (and CR values) created with these matrices are shown in Appendix D. These weight coefficients and importance ranking are visualized in Figure 2.

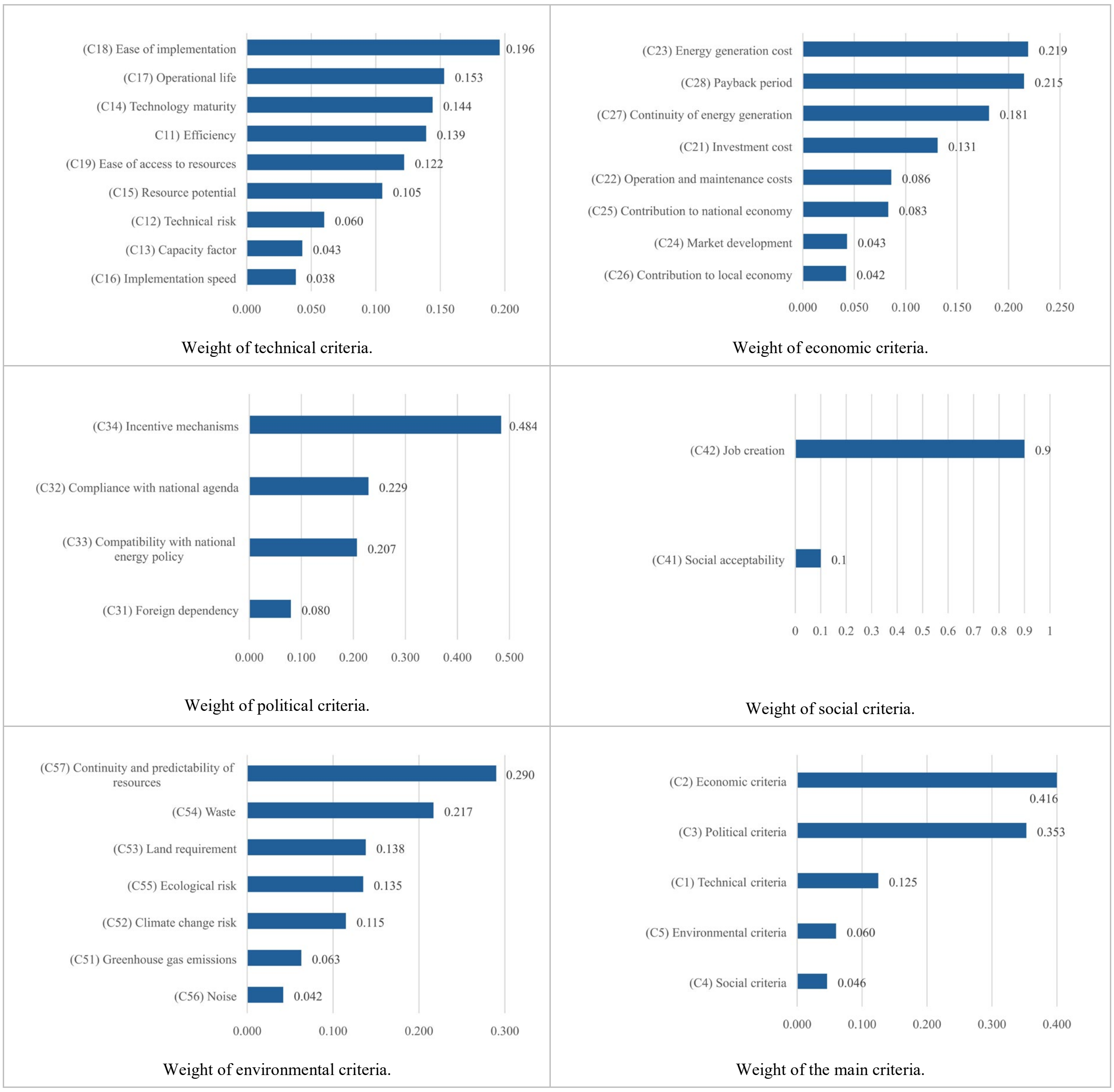

**Figure 2.** Weight of main criteria and sub-criteria

Figure 2 shows the weight of main criteria and sub-criteria. In our study, economic criteria (C2) has the highest weight among the criteria with 41.6%. Economic criteria are followed by political (35.3%), technical (12.5%), environmental (6%) and social (4.6%) criteria. In energy investments, in addition to cost items such as investment cost, operation and maintenance cost, other economic indicators such as contribution to the economy and return on investment are also very important. Therefore, it is appropriate that economic criteria have the highest weight. The importance of environmental factors such as greenhouse gas emissions, climate change, solid waste problems and impacts on the ecosystem in energy investments is indisputable. However, the results of our study show that the importance of environmental aspects in energy investments in Türkiye remains low. The fact that social and environmental dimensions have a lower weight than other main criteria is due to the prioritization of economic and political factors in energy investments in Türkiye. The results of the study are similar to [73], [74].

The analysis results show that the three most important economic sub-criteria are energy generation cost (C23), payback period (C28) and continuity of energy generation (C27). The weights of these criteria are 21.9%, 21.5% and 18.1% respectively. The low cost of electricity generation and the continuity of energy generation are of great importance in the economic evaluation of investments. The short payback period of the investment is also an attractive factor for investors. The close weights between these criteria indicate that the cost of electricity generation and the payback period of the investment are more decisive than the other sub-criteria. The continuity of energy production is directly related to the sustainable use of resources and supports the long-term success of investments. Among the political criteria, the three criteria with the highest weights are incentive mechanisms (C34), compliance with national agenda (C32) and compatibility with national energy policy (C33). The weights of these criteria are 48.4%, 22.9% and 20.7% respectively. Energy investments, which require high capital, need to be supported by government incentives and policies. These supports are critical for attracting investors to the energy sector. While the compatibility of energy investments with political and legal frameworks increases the success of projects, policies that encourage the use of domestic resources to reduce foreign dependency are also important. Foreign dependency (C31) is the last sub-criterion of political criteria with a weight of 8%. In order to reduce foreign dependency in energy, it is important to turn to domestic energy resources. However, in such a case, importing technology will cause the content of the import item to change, while foreign dependency will remain unchanged. Therefore, in order to reduce foreign dependency in energy, technology must not be imported and must be developed domestically.

The highest weights among the technical sub-criteria are ease of implementation (C18), operational life (C17) and technology maturity (C14). We calculated the weights of these sub-criteria as 19.6%, 15.3% and 14.4%, respectively. The ease of implementation and long operational life of power plants ensure that the power plant operates for many years and generates more revenue. High technology maturity significantly helps to reduce foreign dependency in energy investments. On the other hand, efficiency (C11) ranks fourth with a weight of 13.9%. High efficiency is a critical factor desired in energy investments as it directly affects profitability. Regarding environmental criteria, the three criteria with the highest weights are continuity and predictability of resources (C57), waste (C54) and land requirement (C53). We calculated the weights of these sub-criteria as 29.0%, 21.7% and 13.8% respectively. Stable production of power plants is directly related to the sustainability of resources. In addition, waste management is a major cost and environmental impact, especially in geothermal plants, while plants that require little land become more attractive by reducing investment costs. Noise (C56) is the lowest weighted sub-criterion in this category with 4.2%. Noise is particularly important for wind power plants. This is because the mechanical and aerodynamic noise generated by wind power plants can create disturbance in areas close to residential areas.

In the social criteria, job creation (C42) is the most prominent sub-criterion of this category with a weight of 9.0%. In developing countries such as Türkiye, social and environmental aspects are often put on the back agenda, while economic factors become more important. In this regard, employment opportunities provided by energy projects bring job cretion to the forefront. Because the energy sector creates direct and indirect employment and offers job opportunities in many fields such as construction, maintenance, repair and management. On the other hand, social acceptability (C41) was found to be the least important criterion in this category with a weight of 1.0%. This suggests that social and environmental concerns are often considered secondary to economic gains. The social acceptability of energy projects is often balanced against their economic returns.

### 4.2. Prioritization of Renewable Energy Alternatives

At this stage of our study, the weights of energy resources by sub-criteria were calculated and prioritized. The weight coefficients of energy sources according to sub-criteria are given in Appendix D. Figures derived from Appendix D are presented below to make the results more understandable and easier to compare.

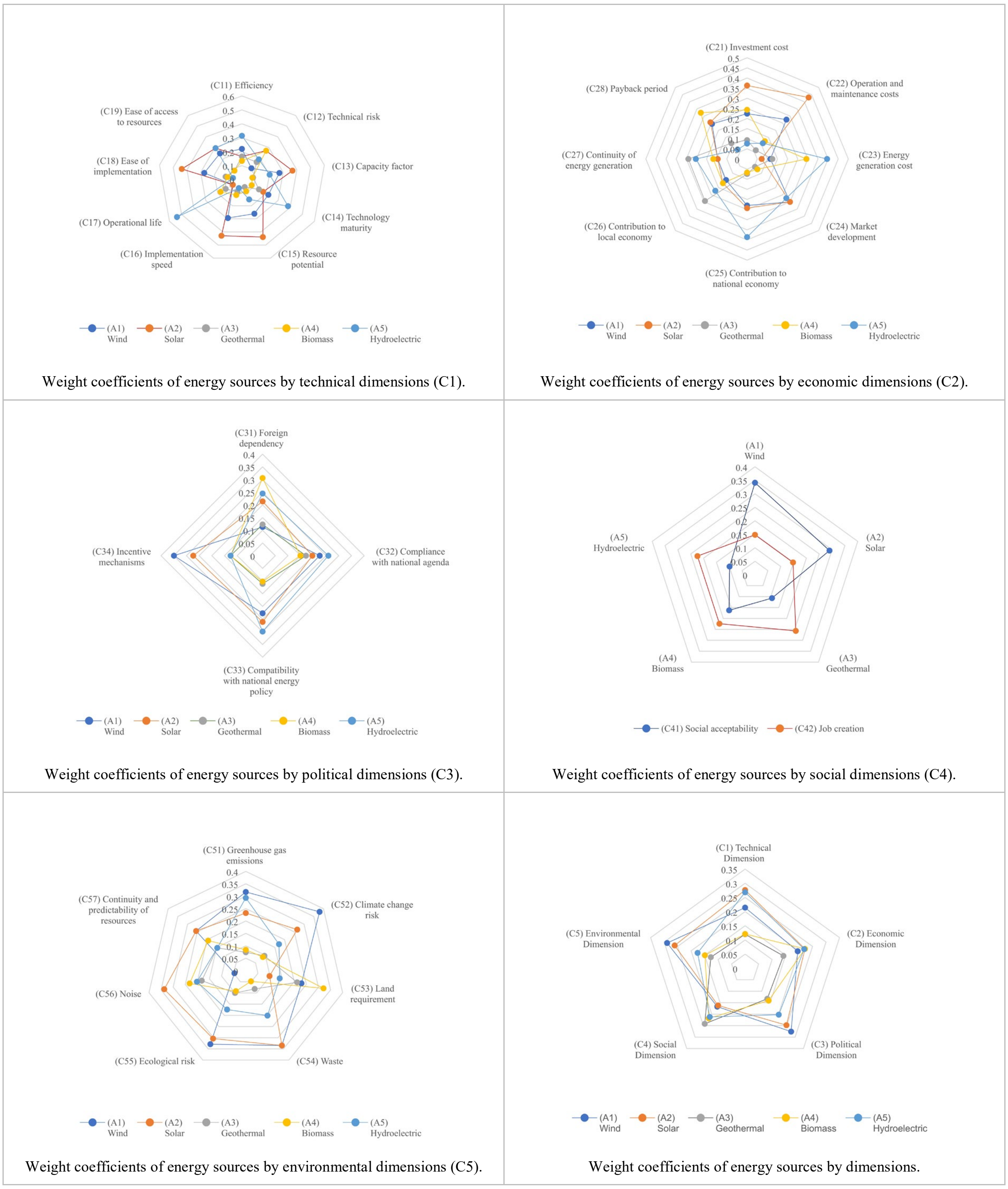

Weight coefficients of energy sources by technical dimensions (C1).

Weight coefficients of energy sources by economic dimensions (C2).

Weight coefficients of energy sources by political dimensions (C3).

Weight coefficients of energy sources by social dimensions (C4).

Weight coefficients of energy sources by environmental dimensions (C5).

Weight coefficients of energy sources by dimensions.

**Figure 3.** Renewable energy's score by main criteria and sub-criteria

Figure 3 shows the score of renewable energy according to the main criteria and sub-criteria. We analyzed each renewable energy source according to both main criteria and sub-criteria. The results of the study show that solar energy (A2) has the highest weighting based on technical (K1) and economic criteria (K2); wind has the highest weighting based on political (K3) and environmental criteria (K5); geothermal energy has the highest weighting based on social criteria (K4).

Wind energy offers significant opportunities in terms of market development and incentive mechanisms. Environmentally, wind turbines have one of the lowest carbon footprints and offer great efficiency in energy production. However, noise and aesthetic concerns during the rotation of the turbines can challenge social acceptance. With careful planning and site selection, the installation of wind turbines can minimize environmental and social impacts. Specifically, areas away from settlements and less ecologically sensitive areas should be preferred. The negative impacts of wind energy on bird migration routes are among the factors that need to be carefully managed. This energy source offers a sustainable and economical energy alternative when supported by financial incentives at local and national level.

Solar energy is technically advantageous in terms of ease of implementation, implementation speed and resource potential. Türkiye's high solar energy potential enhances this advantage. Economically, investment cost, operation and maintenance costs and payback period bring solar energy to the forefront. Solar power plants are also advantageous in terms of market development and are supported by extensive financial support. Socially, they are highly socially accepted due to low environmental damage and local employment opportunities. Environmentally, it stands out as one of the cleanest energy sources with minimal impact on climate change and greenhouse gas emissions. However, extensive land use can be problematic, especially in areas that are valuable for agriculture. The lands selected for the installation of solar panel plants should have sufficient solar radiation as well as low ecological and agricultural value.

Geothermal energy is advantageous in terms of contribution to local economy and continuity of energy generation. Sustainable utilization of underground resources makes geothermal energy particularly attractive for local development. Economically, although it has high investment costs, it has the advantage of continuous and reliable energy generation. Geothermal power plants are among the technologies that minimize environmental damage in social and environmental criteria. The process of treating the water extracted from the ground and injecting it back into the ground ensures the preservation of the environmental balance. However, this process may cause additional costs. It requires transparent management and communication strategies to gain the support of local communities.

Biomass energy plays a critical role in energy security by providing a continuous and predictable energy flow from a technical perspective. However, it is less developed in terms of technology maturity and efficiency compared to other renewable energy sources. Economically, it is the energy source with the shortest payback period since local materials are used. While initial investment costs can be relatively low, operation and maintenance costs can vary depending on the type of biomass used. Political criteria show that biomass energy is generally compatible with local and national energy policies. However, it makes a limited contribution to reducing foreign dependency. It may therefore be a less preferred option for countries with high energy imports. On the environmental dimension, biomass energy offers an environmentally friendly alternative with sustainable resource utilization and waste reduction potential. However, it can produce carbon emissions and other air pollutants during the combustion process. This requires environmental regulation and management. Solid waste and the need for treatment is another environmental issue that needs to be managed, especially in large-scale biomass plants. On the social dimension, biomass plants can provide local employment, creating economic opportunities, especially in rural areas.

Hydroelectric energy offers many technical and economic advantages. Technically, it is characterized by efficiency in power generation, technology maturity and high operational life. The long operational life of the plants, continuous energy supply and generally low operating costs make hydroelectric power plants attractive. However, the construction of these plants requires high initial investments and construction periods are very long. Economically, despite its high initial cost, it stands out with its low electricity generation in the long term and its contribution to the national economy. On the political dimension, hydropower is in line with national energy policies and is supported by long-term financial support and incentives. However, hydropower with low social acceptability can have negative impacts on local communities. Especially during the construction of dams, forced migration and flooded ecosystems bring social and environmental controversies. Environmentally, there are also disadvantages such as evaporation and negative impacts on local flora.

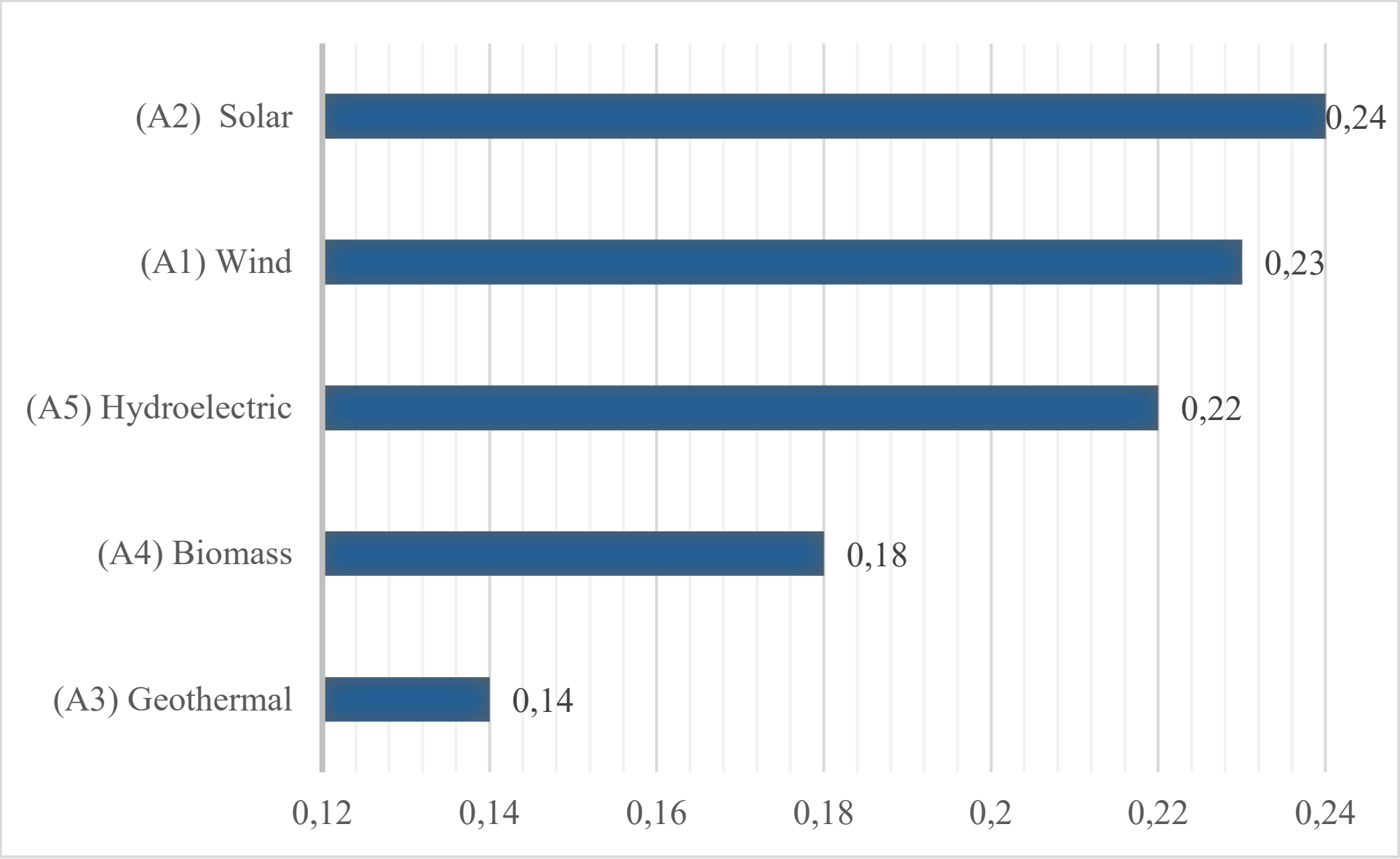

**Figure 4.** Final ranking of the renewable energy alternatives

The ranking of renewable energy sources in our study is solar, wind, hydroelectric, biomass and geothermal. The results of the analysis are consistent with the literature (e.g. [18], [20], [26], [27], [42], [57]). As a result, it was concluded that solar energy is the most suitable renewable energy source to invest in for sustainable development in Türkiye, while wind is the second most suitable renewable energy source.

### 4.3. Sensitivity Analysis

The study conducted a sensitivity analysis by systematically changing the weight values of the main criteria. Therefore, the impacts of the weight changes on the ranking results were analyzed. As a result of the analysis, it is observed how the changes in the criteria weights create a difference in the final ranking.While changing the weights of the main criteria in the sensitivity analysis, we follow the method of Tasri and Susilawati [50]. We set six scenarios to change the weights. Scenario 1 is the original weights in this paper. In Scenario 2, the weight of main criterion C1 is increased by 50%, in Scenario 3 the weight of main criterion C2 is increased by 50% and the other weights are changed proportionally. We followed the same steps for the other scenarios and the results are shown in Figure 5.

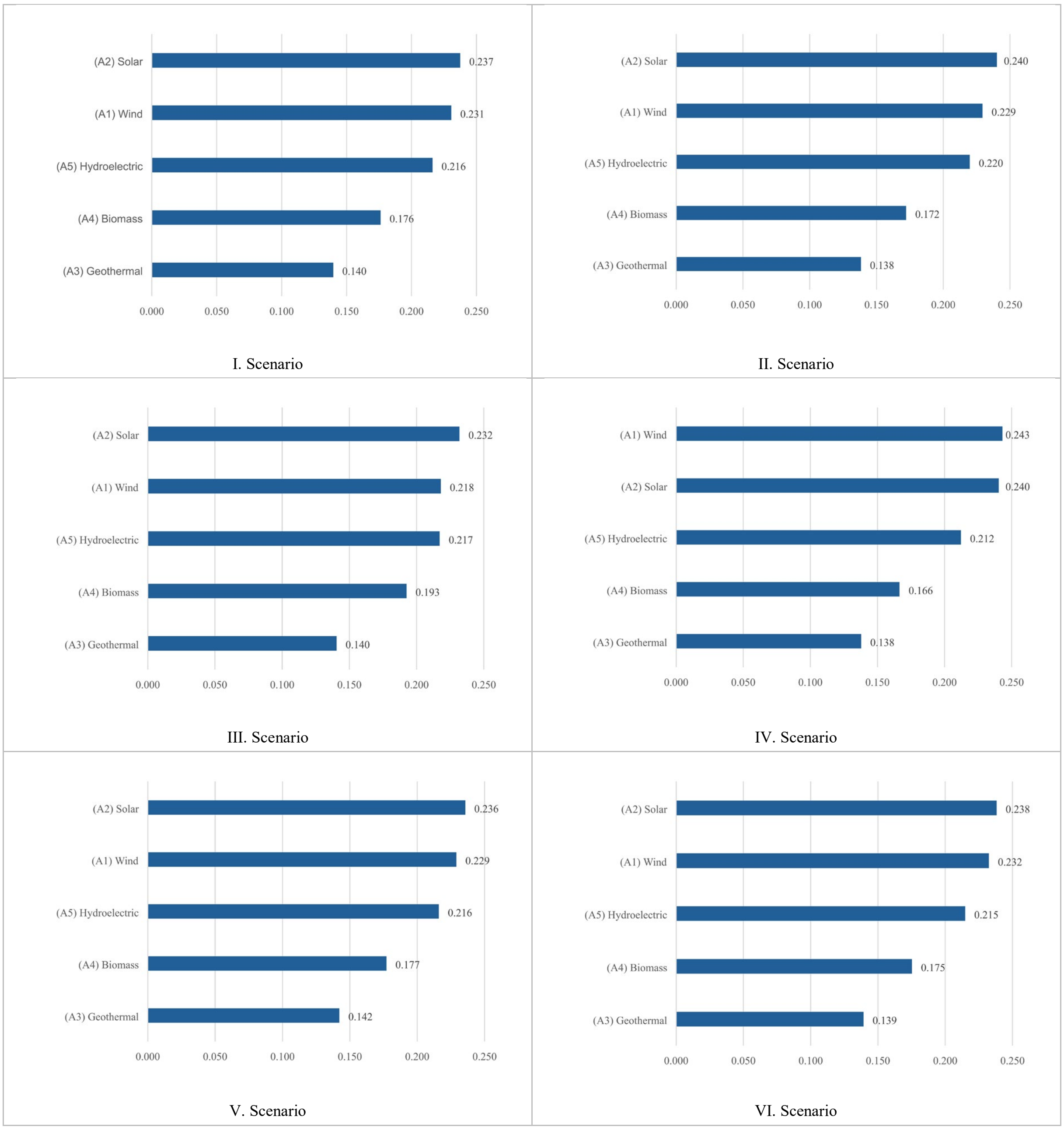

**Figure 5.** Results of the sensitivity analysis

The results of Table 5 show that, except for Scenario 4, changing the weights of the main criteria does not have a significant impact on the ranking of the alternatives. However, in Scenario 4, it is observed that the ranking of solar and wind energy alternatives changes. Overall, the sensitivity analysis demonstrates that there is no significant change in the main findings of the study. Therefore, the impact of changing the weights on the results is insignificant. Solar energy ranks first, followed by wind, hydropower, biomass and geothermal energy. These results show that the study is reliable and consistent.

## 5. POLICY IMPLICATIONS

Our study aims to prioritize renewable energy sources to achieve sustainable development in Türkiye. The results show that solar energy is the most suitable renewable energy source to invest in for sustainable development in Türkiye, followed by wind energy. Türkiye's high solar and wind energy potential provides a great opportunity for sustainable development goals. These results emphasize that solar and wind energy should be considered when designing energy policies in Türkiye. At this stage, policy makers in Türkiye have some important responsibilities. Therefore, various policies have been proposed: (1) Incentives for renewable energy sources, particularly solar and wind, should be increased. Since Türkiye has high solar and wind energy potential, incentives for electricity generation from these sources should be increased. Financial incentives such as tax reductions, low-interest loans and investment subsidies can be applied. (2) Connectivity infrastructure for renewable energy plants needs to be improved. Transmission and distribution networks should be strengthened, and the integration of energy storage systems should be increased. (3) The establishment of hybrid systems should be encouraged to increase the efficiency of variable energy sources such as solar and wind energy. Hybrid systems ensure continuity in energy production and more efficient use of energy resources. (4) R&D investments should be increased to increase the domestic production capacity of renewable energy technologies and ensure technological independence. In this way, technological innovations are encouraged, and local employment is supported. (5) Environmental and social aspects should be given more importance in energy policies. Regulations in this area should be increased. In particular, environmental impact assessments and cooperation with local communities are critical in this process. (6) Training programs and campaigns should be organized to raise public awareness on renewable energy technologies and sustainable energy use. (7) International support and investment should be attracted to renewable energy projects through cooperation on international funding, technology transfer and expert exchange. Implementation of these policies will help Türkiye achieve its sustainable development goals. Implementation of these recommendations requires strong political will and cooperation.

## 6. CONCLUSIONS

In this study, renewable energy sources that can be the best option for sustainable development in Türkiye are analyzed. The results show that the ranking of the main criteria for renewable energy investments in Türkiye is economic, political, technical, environmental and social.

While economic and political criteria are considered more important and ranked first, environmental and social criteria remain in the background. On the basis of economic criteria, solar and biomass energy are the most advantageous renewable sources. With respect to political criteria, wind and solar energy are the best alternatives. Technically, solar and hydroelectric energy are more suitable. Solar energy is the best option when environmental criteria are considered, while solar and geothermal energy are among the best options in terms of social criteria. Based on all the criteria used in our study, we conclude that solar energy is the most suitable renewable energy source to invest in for sustainable development in Türkiye, while wind is the second most suitable renewable energy source.

This study has some limitations as well as important findings. These limitations can be addressed in future studies. Firstly, this study used 30 criteria under 5 dimensions. The study can be improved by increasing the number of main and sub-criteria. Secondly, renewable energy sources with the most widespread use and the highest potential for Türkiye were preferred in the study. Therefore, renewable energy sources other than wind, solar, geothermal, geothermal, biomass and hydropower were not included in the analysis. Thirdly, the results of the study can be compared by using different and/or hybrid techniques. Finally, this study did not analyze the hybrid use of renewable energy sources. Further research in this area is recommended.

## NOMENCLATURE

| | |
|---|---|
| AHP | : Analytic Hierarchy Process |
| CI | : Consistency Index |
| CODAS | : Combinative Distance-based Assessment |
| CR | : Consistency Ratio |
| DEMATEL | : Decision Making Trial and Evaluation Laboratory |
| GDP | : Gross Domestic Product |
| MACBETH | : Measuring Attractiveness by a Categorical Based Evaluation Technique |
| PROMETHEE | : Preference Ranking Organization Method for Enrichment Evaluations |
| R&D | : Research and Development |
| RI | : Random Index |
| SWARA | : Stepwise Weight Assessment Ratio Analysis |
| TOPSIS | : Technique for Order Preference by Similarity to Ideal Solution |
| WASPAS | : Weighted Aggregated Sum Product Assessment |

## ACKNOWLEDGMENT

This research is not supported by any public or commercial institution or organization.

## DECLARATION OF ETHICAL STANDARDS

The authors of the paper submitted declare that nothing which is necessary for achieving the paper requires ethical committee and/or legal-special permissions.

## CONTRIBUTION OF THE AUTHORS

**Emre Akusta:** Writing, Investigation, Analysis, Methodology, Conceptualization.

**Raif Cergibozan:** Conceptualization, Methodology, Editing, Supervision, Resources.

## CONFLICT OF INTEREST

There is no conflict of interest in this study.

## APPENDIX A

**Table A.1.** Extended version of the criteria used in the study with references

| Criteria | Sub-criteria | | Reference |
|---|---|---|---|
| $C_1$. Technical | $C_{11}$ | Efficiency | [23], [24], [28], [29], [32], [33], [35], [36], [37], [38], [40], [41], [76], [77], [78], [79], [80], [81], [82] |
| | $C_{12}$ | Technical risk | [18], [23], [24], [25], [28], [35], [37], [40], [41], [77], [79], [80], [81], [83], [84], [85] |
| | $C_{13}$ | Capacity factor | [27], [28], [33], [35], [38], [41], [77], [78], [79], [80], [81] |
| | $C_{14}$ | Technology maturity | [20], [21], [24], [27], [28], [30], [31], [32], [35], [37], [38], [41], [77], [80], [81], [82], [84] |
| | $C_{15}$ | Resource potential | [25], [27], [31], [33], [35], [41], [78], [79], [82] |
| | $C_{16}$ | Implementation speed | [28], [29], [35], [37], [40], [41], [79], [81] |
| | $C_{17}$ | Operational life | [23], [28], [29], [33], [35], [37], [40], [41], [78], [79], [80], [81], [82], [84] |
| | $C_{18}$ | Ease of implementation | [24], [28], [37], [38], [40], [41] |
| | $C_{19}$ | Ease of access to resources | [24], [28], [35], [40], [41], [85] |
| $C_2$. Economic | $C_{21}$ | Investment cost | [27], [29], [31], [33], [35], [36], [37], [40], [41], [76], [77], [78], [79], [80], [83], [84], [86] |
| | $C_{22}$ | Operation and maintenance costs | [27], [28], [29], [31], [32], [33], [35], [36], [37], [40], [41], [76], [78], [79], [80], [81], [83], [86] |
| | $C_{23}$ | Energy generation cost | [23], [27], [28], [31], [32], [35], [36], [38], [40], [41], [78], [79], [80], [81], [82] |
| | $C_{24}$ | Market development | [21], [41], [78] |
| | $C_{25}$ | Contribution to national economy | [28], [41], [77], [80] |
| | $C_{26}$ | Contribution to local economy | [21], [41], [80], [81], [82] |
| | $C_{27}$ | Continuity of energy generation | [24], [27], [28], [31], [32], [41] |
| | $C_{28}$ | Payback period | [23], [27], [28], [33], [41], [78], [79], [80], [81], [82] |
| $C_3$. Political | $C_{31}$ | Foreign dependency | [21], [27], [29], [35], [37], [38] |
| | $C_{32}$ | Compliance with national agenda | [27], [31], [35], [41], [79], [81], [84] |
| | $C_{33}$ | Compatibility with national energy policy | [21], [27], [28], [32], [35], [40], [41], [79], [80], [81] |
| | $C_{34}$ | Incentive mechanisms | [21], [27], [28], [29], [31], [32], [35], [37], [40], [41], [77], [79], [81], [82] |
| $C_4$. Social | $C_{41}$ | Social acceptability | [17], [18], [20], [23], [28], [29], [32], [35], [36], [37], [40], [77], [80], [84] |
| | $C_{42}$ | Job creation | [18], [22], [23], [23], [27], [28], [29], [31], [32], [33], [35], [36], [37], [38], [40], [41], [76], [79], [82] |
| $C_5$. Environmental | $C_{51}$ | Greenhouse gas emissions | [18], [21], [24], [29], [31], [33], [35], [37], [40], [41], [77], [78], [79], [80], [81], [82], [84], [87] |
| | $C_{52}$ | Climate change risk | [18], [21], [24], [25], [27], [28], [33], [35], [38], [40], [41], [78], [79], [80], [81], [84], [87] |
| | $C_{53}$ | Land requirement | [18], [21], [23], [28], [29], [33], [35], [36], [37], [40], [41], [76], [77], [79], [80], [82], [84] |
| | $C_{54}$ | Waste | [18], [21], [24], [28], [35], [40], [41], [77], [79], [80], [81], [84] |
| | $C_{55}$ | Ecological risk | [18], [23], [24], [25], [27], [28], [29], [31], [32], [33], [35], [37], [41], [77], [78], [84], [87] |
| | $C_{56}$ | Noise | [24], [28], [37], [41], [76], [78], [81] |
| | $C_{57}$ | Continuity and predictability of resources | [18], [28], [30], [35], [37], [41], [85], [88] |

## APPENDIX B

**Table A.1.** Profiles of experts

| S. No. | Position of expert | Age | Qualification | Department/Organization |
|---|---|---|---|---|
| 1 | Professor | 40 - 50 | Ph.D | Academia (University) |
| 2 | Professor | 30 - 40 | Ph.D | Academia (University) |
| 3 | Assoc. Prof | 30 - 40 | Ph.D | Academia (University) |
| 4 | Assoc. Prof | 30 - 40 | Ph.D | Academia (University) |
| 5 | Assoc. Prof | 40 - 50 | Ph.D | Academia (University) |
| 6 | Assoc. Prof | 30 - 40 | Ph.D | Academia (University) |
| 7 | Assoc. Prof | 30 - 40 | Ph.D | Academia (University) |
| 8 | Assist. Prof | 30 - 40 | Ph.D | Academia (University) |
| 9 | Assist. Prof | 30 - 40 | Ph.D | Academia (University) |
| 10 | Assist. Prof | 20 - 30 | Ph.D | Academia (University) |
| 11 | Assist. Prof | 20 - 30 | Ph.D | Academia (University) |
| 12 | Assist. Prof | 20 - 30 | Ph.D | Academia (University) |
| 13 | Assist. Prof | 20 - 30 | Ph.D | Academia (University) |

## APPENDIX C

**Table B.1.** Main criteria matrix.

| | $C_1$ | | | $C_2$ | | | $C_3$ | | | $C_4$ | | | $C_5$ | | |
|---|---|---|---|---|---|---|---|---|---|---|---|---|---|---|---|
| **$C_1$** | 1.00 | 1.00 | 1.00 | 0.17 | 0.20 | 0.25 | 0.25 | 0.33 | 0.50 | 2.00 | 3.00 | 4.00 | 2.00 | 3.00 | 4.00 |
| **$C_2$** | 4.00 | 5.00 | 6.00 | 1.00 | 1.00 | 1.00 | 1.00 | 1.00 | 1.00 | 6.00 | 7.00 | 8.00 | 6.00 | 7.00 | 8.00 |
| **$C_3$** | 2.00 | 3.00 | 4.00 | 1.00 | 1.00 | 1.00 | 1.00 | 1.00 | 1.00 | 5.00 | 6.00 | 7.00 | 5.00 | 6.00 | 7.00 |
| **$C_4$** | 0.25 | 0.33 | 0.50 | 0.13 | 0.14 | 0.17 | 0.14 | 0.17 | 0.20 | 1.00 | 1.00 | 1.00 | 0.33 | 0.50 | 1.00 |
| **$C_5$** | 0.25 | 0.33 | 0.50 | 0.13 | 0.14 | 0.17 | 0.14 | 0.17 | 0.20 | 1.00 | 2.00 | 3.00 | 1.00 | 1.00 | 1.00 |
| CR=0.087 | | | | | | | | | | | | | | | |

**Table B.2.** Technical criteria matrix.

| | $C_{11}$ | | | $C_{12}$ | | | $C_{13}$ | | | $C_{14}$ | | | $C_{15}$ | | | $C_{16}$ | | | $C_{17}$ | | | $C_{18}$ | | | $C_{19}$ | | |
|---|---|---|---|---|---|---|---|---|---|---|---|---|---|---|---|---|---|---|---|---|---|---|---|---|---|---|---|
| **$C_{11}$** | 1.00 | 1.00 | 1.00 | 3.00 | 4.00 | 5.00 | 4.00 | 5.00 | 6.00 | 0.33 | 0.50 | 1.00 | 1.00 | 1.00 | 1.00 | 4.00 | 5.00 | 6.00 | 1.00 | 1.00 | 1.00 | 1.00 | 1.00 | 1.00 | 0.33 | 0.50 | 1.00 |
| **$C_{12}$** | 0.20 | 0.25 | 0.33 | 1.00 | 1.00 | 1.00 | 1.00 | 2.00 | 3.00 | 0.33 | 0.50 | 1.00 | 0.33 | 0.50 | 1.00 | 1.00 | 1.00 | 1.00 | 0.25 | 0.33 | 0.50 | 0.25 | 0.33 | 0.50 | 1.00 | 1.00 | 1.00 |
| **$C_{13}$** | 0.17 | 0.20 | 0.25 | 0.33 | 0.50 | 1.00 | 1.00 | 1.00 | 1.00 | 0.33 | 0.50 | 1.00 | 0.33 | 0.50 | 1.00 | 1.00 | 1.00 | 1.00 | 0.20 | 0.25 | 0.33 | 0.25 | 0.33 | 0.50 | 0.25 | 0.33 | 0.50 |
| **$C_{14}$** | 1.00 | 2.00 | 3.00 | 1.00 | 2.00 | 3.00 | 1.00 | 2.00 | 3.00 | 1.00 | 1.00 | 1.00 | 1.00 | 2.00 | 3.00 | 2.00 | 3.00 | 4.00 | 0.33 | 0.50 | 1.00 | 0.33 | 0.50 | 1.00 | 2.00 | 3.00 | 4.00 |
| **$C_{15}$** | 1.00 | 1.00 | 1.00 | 1.00 | 2.00 | 3.00 | 1.00 | 2.00 | 3.00 | 0.33 | 0.50 | 1.00 | 1.00 | 1.00 | 1.00 | 3.00 | 4.00 | 5.00 | 1.00 | 1.00 | 1.00 | 0.33 | 0.50 | 1.00 | 0.33 | 0.50 | 1.00 |
| **$C_{16}$** | 0.17 | 0.20 | 0.25 | 1.00 | 1.00 | 1.00 | 1.00 | 1.00 | 1.00 | 0.25 | 0.33 | 0.50 | 0.20 | 0.25 | 0.33 | 1.00 | 1.00 | 1.00 | 0.17 | 0.20 | 0.25 | 0.20 | 0.25 | 0.33 | 0.20 | 0.25 | 0.33 |
| **$C_{17}$** | 1.00 | 1.00 | 1.00 | 2.00 | 3.00 | 4.00 | 3.00 | 4.00 | 5.00 | 1.00 | 2.00 | 3.00 | 1.00 | 1.00 | 1.00 | 4.00 | 5.00 | 6.00 | 1.00 | 1.00 | 1.00 | 0.33 | 0.50 | 1.00 | 1.00 | 1.00 | 1.00 |
| **$C_{18}$** | 1.00 | 1.00 | 1.00 | 2.00 | 3.00 | 4.00 | 2.00 | 3.00 | 4.00 | 1.00 | 2.00 | 3.00 | 1.00 | 2.00 | 3.00 | 3.00 | 4.00 | 5.00 | 1.00 | 2.00 | 3.00 | 1.00 | 1.00 | 1.00 | 1.00 | 2.00 | 3.00 |
| **$C_{19}$** | 1.00 | 2.00 | 3.00 | 1.00 | 1.00 | 1.00 | 2.00 | 3.00 | 4.00 | 0.25 | 0.33 | 0.50 | 1.00 | 2.00 | 3.00 | 3.00 | 4.00 | 5.00 | 1.00 | 1.00 | 1.00 | 0.33 | 0.50 | 1.00 | 1.00 | 1.00 | 1.00 |
| CR=0.058 | | | | | | | | | | | | | | | | | | | | | | | | | | | |

**Table B.3.** Economic criteria matrix.

| | $C_{21}$ | | | $C_{22}$ | | | $C_{23}$ | | | $C_{24}$ | | | $C_{25}$ | | | $C_{26}$ | | | $C_{27}$ | | | $C_{28}$ | | |
|---|---|---|---|---|---|---|---|---|---|---|---|---|---|---|---|---|---|---|---|---|---|---|---|---|
| $C_{21}$ | 1.00 | 1.00 | 1.00 | 2.00 | 3.00 | 4.00 | 0.33 | 0.50 | 1.00 | 3.00 | 4.00 | 5.00 | 1.00 | 1.00 | 1.00 | 3.00 | 4.00 | 5.00 | 0.33 | 0.50 | 1.00 | 0.33 | 0.50 | 1.00 |
| $C_{22}$ | 0.25 | 0.33 | 0.50 | 1.00 | 1.00 | 1.00 | 0.25 | 0.33 | 0.50 | 4.00 | 5.00 | 6.00 | 1.00 | 1.00 | 1.00 | 2.00 | 3.00 | 4.00 | 0.33 | 0.50 | 1.00 | 0.20 | 0.25 | 0.33 |
| $C_{23}$ | 1.00 | 2.00 | 3.00 | 2.00 | 3.00 | 4.00 | 1.00 | 1.00 | 1.00 | 4.00 | 5.00 | 6.00 | 4.00 | 5.00 | 6.00 | 4.00 | 5.00 | 6.00 | 0.33 | 0.50 | 1.00 | 1.00 | 1.00 | 1.00 |
| $C_{24}$ | 0.20 | 0.25 | 0.33 | 0.17 | 0.20 | 0.25 | 0.17 | 0.20 | 0.25 | 1.00 | 1.00 | 1.00 | 1.00 | 1.00 | 1.00 | 1.00 | 2.00 | 3.00 | 0.17 | 0.20 | 0.25 | 0.17 | 0.20 | 0.25 |
| $C_{25}$ | 1.00 | 1.00 | 1.00 | 1.00 | 1.00 | 1.00 | 0.17 | 0.20 | 0.25 | 1.00 | 1.00 | 1.00 | 1.00 | 1.00 | 1.00 | 2.00 | 3.00 | 4.00 | 0.33 | 0.50 | 1.00 | 0.33 | 0.50 | 1.00 |
| $C_{26}$ | 0.20 | 0.25 | 0.33 | 0.25 | 0.33 | 0.50 | 0.17 | 0.20 | 0.25 | 0.33 | 0.50 | 1.00 | 0.25 | 0.33 | 0.50 | 1.00 | 1.00 | 1.00 | 0.33 | 0.50 | 1.00 | 0.33 | 0.50 | 1.00 |
| $C_{27}$ | 1.00 | 2.00 | 3.00 | 1.00 | 2.00 | 3.00 | 1.00 | 2.00 | 3.00 | 4.00 | 5.00 | 6.00 | 1.00 | 2.00 | 3.00 | 1.00 | 2.00 | 3.00 | 1.00 | 1.00 | 1.00 | 0.33 | 0.50 | 1.00 |
| $C_{28}$ | 1.00 | 2.00 | 3.00 | 3.00 | 4.00 | 5.00 | 1.00 | 1.00 | 1.00 | 4.00 | 5.00 | 6.00 | 1.00 | 2.00 | 3.00 | 1.00 | 2.00 | 3.00 | 1.00 | 2.00 | 3.00 | 1.00 | 1.00 | 1.00 |
| CR=0.099 | | | | | | | | | | | | | | | | | | | | | | | | |

**Table B.4.** Political criteria matrix.

| | $C_{31}$ | | | $C_{32}$ | | | $C_{33}$ | | | $C_{34}$ | | |
|---|---|---|---|---|---|---|---|---|---|---|---|---|
| $C_{31}$ | 1.00 | 1.00 | 1.00 | 0.25 | 0.33 | 0.50 | 0.25 | 0.33 | 0.50 | 0.17 | 0.20 | 0.25 |
| $C_{32}$ | 2.00 | 3.00 | 4.00 | 1.00 | 1.00 | 1.00 | 1.00 | 1.00 | 1.00 | 0.33 | 0.50 | 1.00 |
| $C_{33}$ | 2.00 | 3.00 | 4.00 | 1.00 | 1.00 | 1.00 | 1.00 | 1.00 | 1.00 | 0.25 | 0.33 | 0.50 |
| $C_{34}$ | 4.00 | 5.00 | 6.00 | 1.00 | 2.00 | 3.00 | 2.00 | 3.00 | 4.00 | 1.00 | 1.00 | 1.00 |
| CR=0.086 | | | | | | | | | | | | |

**Table B.5.** Social criteria matrix.

| | $C_{41}$ | | | $C_{42}$ | | |
|---|---|---|---|---|---|---|
| $C_{41}$ | 1.00 | 1.00 | 1.00 | 0.11 | 0.11 | 0.13 |
| $C_{42}$ | 8.00 | 9.00 | 9.00 | 1.00 | 1.00 | 1.00 |
| CR=0.007 | | | | | | |

**Table B.6.** Environmental criteria matrix.

| | $C_{51}$ | | | $C_{52}$ | | | $C_{53}$ | | | $C_{54}$ | | | $C_{55}$ | | | $C_{56}$ | | | $C_{57}$ | | |
|---|---|---|---|---|---|---|---|---|---|---|---|---|---|---|---|---|---|---|---|---|---|
| $C_{51}$ | 1.00 | 1.00 | 1.00 | 1.00 | 1.00 | 1.00 | 0.20 | 0.25 | 0.33 | 0.20 | 0.25 | 0.33 | 0.25 | 0.33 | 0.50 | 1.00 | 2.00 | 3.00 | 0.20 | 0.25 | 0.33 |
| $C_{52}$ | 1.00 | 1.00 | 1.00 | 1.00 | 1.00 | 1.00 | 1.00 | 1.00 | 1.00 | 0.33 | 0.50 | 1.00 | 1.00 | 1.00 | 1.00 | 2.00 | 3.00 | 4.00 | 0.33 | 0.50 | 1.00 |
| $C_{53}$ | 3.00 | 4.00 | 5.00 | 1.00 | 1.00 | 1.00 | 1.00 | 1.00 | 1.00 | 0.33 | 0.50 | 1.00 | 1.00 | 1.00 | 1.00 | 3.00 | 4.00 | 5.00 | 0.25 | 0.33 | 0.50 |
| $C_{54}$ | 3.00 | 4.00 | 5.00 | 1.00 | 2.00 | 3.00 | 1.00 | 2.00 | 3.00 | 1.00 | 1.00 | 1.00 | 1.00 | 2.00 | 3.00 | 3.00 | 4.00 | 5.00 | 0.33 | 0.50 | 1.00 |
| $C_{55}$ | 2.00 | 3.00 | 4.00 | 1.00 | 1.00 | 1.00 | 1.00 | 1.00 | 1.00 | 0.33 | 0.50 | 1.00 | 1.00 | 1.00 | 1.00 | 2.00 | 3.00 | 4.00 | 0.33 | 0.50 | 1.00 |
| $C_{56}$ | 0.33 | 0.50 | 1.00 | 0.25 | 0.33 | 0.50 | 0.20 | 0.25 | 0.33 | 0.20 | 0.25 | 0.33 | 0.25 | 0.33 | 0.50 | 1.00 | 1.00 | 1.00 | 0.17 | 0.20 | 0.25 |
| $C_{57}$ | 3.00 | 4.00 | 5.00 | 1.00 | 2.00 | 3.00 | 2.00 | 3.00 | 4.00 | 1.00 | 2.00 | 3.00 | 1.00 | 2.00 | 3.00 | 4.00 | 5.00 | 6.00 | 1.00 | 1.00 | 1.00 |
| CR=0.077 | | | | | | | | | | | | | | | | | | | | | |

**APPENDIX D**

**Table C.1.** Weights of renewable energy sources by technical criteria (C1).

| | $C_{11}$ | $C_{12}$ | $C_{13}$ | $C_{14}$ | $C_{15}$ | $C_{16}$ | $C_{17}$ | $C_{18}$ | $C_{19}$ | R.E.S. weights |
|---|---|---|---|---|---|---|---|---|---|---|
| **$A_1$** | 0.22 | 0.106 | 0.273 | 0.219 | 0.261 | 0.294 | 0.075 | 0.274 | 0.245 | 0.214 |
| **$A_2$** | 0.166 | 0.27 | 0.368 | 0.176 | 0.439 | 0.429 | 0.075 | 0.438 | 0.294 | 0.276 |
| **$A_3$** | 0.166 | 0.164 | 0.081 | 0.141 | 0.057 | 0.095 | 0.134 | 0.119 | 0.084 | 0.120 |
| **$A_4$** | 0.134 | 0.27 | 0.076 | 0.081 | 0.09 | 0.118 | 0.177 | 0.104 | 0.083 | 0.121 |
| **$A_5$** | 0.314 | 0.189 | 0.202 | 0.382 | 0.152 | 0.065 | 0.538 | 0.065 | 0.294 | 0.268 |
| **M.C. Weights** | 0.139 | 0.06 | 0.043 | 0.144 | 0.105 | 0.038 | 0.153 | 0.196 | 0.122 | |
| **CR** | 0.059 | 0.066 | 0.088 | 0.061 | 0.087 | 0.079 | 0.097 | 0.086 | 0.077 | |

Note: M.C., R.E.S., and CR denote main criteria, renewable energy sources, and Consistency rate, respectively.

**Table C.2 .** Weights of renewable energy sources by economic criteria (C2).

| | $C_{21}$ | $C_{22}$ | $C_{23}$ | $C_{24}$ | $C_{25}$ | $C_{26}$ | $C_{27}$ | $C_{28}$ | R.E.S. weights |
|---|---|---|---|---|---|---|---|---|---|
| **$A_1$** | 0.224 | 0.275 | 0.115 | 0.300 | 0.230 | 0.147 | 0.145 | 0.245 | 0.195 |
| **$A_2$** | 0.363 | 0.430 | 0.072 | 0.300 | 0.243 | 0.169 | 0.145 | 0.256 | 0.222 |
| **$A_3$** | 0.093 | 0.061 | 0.125 | 0.055 | 0.074 | 0.294 | 0.290 | 0.110 | 0.142 |
| **$A_4$** | 0.243 | 0.124 | 0.293 | 0.072 | 0.067 | 0.169 | 0.167 | 0.323 | 0.222 |
| **$A_5$** | 0.076 | 0.110 | 0.395 | 0.274 | 0.386 | 0.222 | 0.253 | 0.066 | 0.219 |
| **M.C. Weights** | 0.131 | 0.086 | 0.219 | 0.043 | 0.083 | 0.042 | 0.181 | 0.215 | |
| **CR** | 0.096 | 0.095 | 0.060 | 0.052 | 0.057 | 0.055 | 0.043 | 0.084 | |

Note: M.C., R.E.S., and CR denote main criteria, renewable energy sources, and Consistency rate, respectively.

**Table C.3.** Weights of renewable energy sources by political criteria (C3).

| | $C_{31}$ | $C_{32}$ | $C_{33}$ | $C_{34}$ | R.E.S. weights |
|---|---|---|---|---|---|
| $A_1$ | 0.113 | 0.225 | 0.227 | 0.349 | 0.276 |
| $A_2$ | 0.214 | 0.196 | 0.261 | 0.273 | 0.248 |
| $A_3$ | 0.123 | 0.171 | 0.111 | 0.126 | 0.133 |
| $A_4$ | 0.306 | 0.149 | 0.102 | 0.126 | 0.141 |
| $A_5$ | 0.245 | 0.259 | 0.299 | 0.126 | 0.202 |
| **M.C. Weights** | 0.080 | 0.229 | 0.207 | 0.484 | |
| **CR** | 0.051 | 0.043 | 0.059 | 0.050 | |

Note: M.C., R.E.S., and CR denote main criteria, renewable energy sources, and Consistency rate, respectively.

**Table C.4.** Weights of renewable energy sources by social criteria (C4).

| | $C_{41}$ | $C_{42}$ | R.E.S. weights |
|---|---|---|---|
| $A_1$ | 0.341 | 0.148 | 0.167 |
| $A_2$ | 0.29 | 0.148 | 0.162 |
| $A_3$ | 0.107 | 0.257 | 0.242 |
| $A_4$ | 0.163 | 0.224 | 0.218 |
| $A_5$ | 0.099 | 0.224 | 0.212 |
| **M.C. Weights** | 0.100 | 0.900 | |
| **CR** | 0.056 | 0.059 | |

Note: M.C., R.E.S., and CR denote main criteria, renewable energy sources, and Consistency rate, respectively.

**Table C.5.** Weights of renewable energy sources by environmental criteria (C5).

| | $C_{51}$ | $C_{52}$ | $C_{53}$ | $C_{54}$ | $C_{55}$ | $C_{56}$ | $C_{57}$ | R.E.S. weights |
|---|---|---|---|---|---|---|---|---|
| $A_1$ | 0.317 | 0.380 | 0.230 | 0.335 | 0.329 | 0.047 | 0.256 | 0.289 |
| $A_2$ | 0.232 | 0.265 | 0.098 | 0.335 | 0.304 | 0.337 | 0.256 | 0.261 |
| $A_3$ | 0.074 | 0.096 | 0.212 | 0.082 | 0.100 | 0.182 | 0.147 | 0.127 |
| $A_4$ | 0.084 | 0.088 | 0.321 | 0.048 | 0.092 | 0.232 | 0.194 | 0.149 |
| $A_5$ | 0.293 | 0.171 | 0.140 | 0.201 | 0.174 | 0.202 | 0.147 | 0.176 |
| **M.C. Weights** | 0.063 | 0.115 | 0.138 | 0.217 | 0.135 | 0.042 | 0.290 | |
| **CR** | 0.090 | 0.074 | 0.067 | 0.046 | 0.066 | 0.071 | 0.065 | |

Note: M.C., R.E.S., and CR denote main criteria, renewable energy sources, and Consistency rate, respectively.

**Table C.6.** Weights of renewable energy sources by all criteria.

| | $C_1$ | $C_2$ | $C_3$ | $C_4$ | $C_5$ | R.E.S. weights |
|---|---|---|---|---|---|---|
| $A_1$ | 0.214 | 0.195 | 0.276 | 0.167 | 0.289 | 0.23 |
| $A_2$ | 0.276 | 0.222 | 0.248 | 0.162 | 0.261 | 0.24 |
| $A_3$ | 0.120 | 0.142 | 0.133 | 0.242 | 0.127 | 0.14 |
| $A_4$ | 0.121 | 0.222 | 0.141 | 0.218 | 0.149 | 0.18 |
| $A_5$ | 0.268 | 0.219 | 0.202 | 0.212 | 0.176 | 0.22 |
| **M.C. Weights** | 0.125 | 0.416 | 0.353 | 0.046 | 0.060 | |

Note: M.C. and R.E.S. denote main criteria and renewable energy sources respectively.

## REFERENCES

[1] Kızıldere C. Türkiye'de cari açık sorununun enerji tüketimi ve ekonomik büyüme açısından değerlendirilmesi: Ampirik bir analiz. Business & Management Studies: An International Journal 2020; 8(2): 2121-2139.

[2] Outlook AE. Annual Energy Outlook 2019: with Projections to 2050. US Energy Information Administration 2019.

[3] Cherp A, Jewell J. The concept of energy security: Beyond the four As. Energy Policy 2014; 75: 415-421.

[4] Rosen MA. Issues, concepts and applications for sustainability. Glocalism: Journal of Culture, Politics and Innovation 2018; 3(1): 5-22.

[5] Tomislav K. The concept of sustainable development: From its beginning to the contemporary issues. Zagreb International Review of Economics & Business 2018; 21(1): 67-94.

[6] Gürlük S. Sürdürülebilir kalkınma gelişmekte olan ülkelerde uygulanabilir mi? Eskişehir Osmangazi Üniversitesi İİBF Dergisi 2010; 5(2): 85-99.

[7] Ibimilua FO. Linkages between poverty and environmental degradation. African Research Review 2011; 5(1): 102-121.

[8] Neumayer E. The human development index and sustainability-a constructive proposal. Ecological Economics 2001; 39(1): 101-114.

[9] Ediger VŞ, Huvaz O. Examining the sectoral energy use in Turkish economy (1980–2000) with the help of decomposition analysis. Energy Conversion and Management 2006; 47(6): 732-745.

[10] Ediger VŞ. Türkiye'nin Sürdürülebilir Enerji Gelişimi. TÜBA, Günce 2009; 39(1): 18-25.

[11] Batı O. Türkiye'de yenilenebilir enerji kaynaklarının sürdürülebilir kalkınmaya etkisi konusunda bir alan araştırması. Trakya Üniversitesi Sosyal Bilimler Dergisi 2014; 16(2): 27-38.

[12] Fotis P, Polemis M. Sustainable development, environmental policy and renewable energy use: A dynamic panel data approach. Sustainable Development 2018; 26(6): 726-740.

[13] Güney T. Renewable energy, non-renewable energy and sustainable development. International Journal of Sustainable Development & World Ecology 2019; 26(5): 389-397.

[14] Dinçer H, Karakuş H. Yenilenebilir enerjinin sürdürülebilir ekonomik kalkınma üzerindeki etkisi: BRICS ve MINT ülkeleri üzerine karşılaştırmalı bir analiz. ESAM Dergisi 2020; 1(1): 75-99.

[15] Öymen G. Yenilenebilir enerjinin sürdürülebilirlik üzerindeki rolü. İstanbul Ticaret Üniversitesi Sosyal Bilimler Dergisi 2020; 19(39): 1069-1087.

[16] Tiba S, Belaid F. Modeling the nexus between sustainable development and renewable energy: The African perspectives. Journal of Economic Surveys 2021; 35(1): 307-329.

[17] Kahraman C, Kaya İ, Cebi S. A comparative analysis for multiattribute selection among renewable energy alternatives using fuzzy axiomatic design and fuzzy analytic hierarchy process. Energy 2009; 34(10): 1603-1616.

[18] Kahraman C, Kaya I. A fuzzy multicriteria methodology for selection among energy alternatives. Expert Systems with Applications 2010; 37(9): 6270-6281.

[19] Atmaca E, Basar HB. Evaluation of power plants in Turkey using Analytic Network Process (ANP). Energy 2012; 44(1): 555-563.

[20] Demirtas O. Evaluating the best renewable energy technology for sustainable energy planning. International Journal of Energy Economics and Policy 2013; 3(4): 23-33.

[21] Yakıcı Ayan T, Pabuçcu H. Yenilenebilir enerji kaynakları yatırım projelerinin analitik hiyerarşi süreci yöntemi ile değerlendirilmesi. Süleyman Demirel Üniversitesi İktisadi ve İdari Bilimler Fakültesi Dergisi 2013; 18(3), 89-110.

[22] Kabak M, Dağdeviren M. Prioritization of renewable energy sources for Turkey by using a hybrid MCDM methodology. Energy Conversion and Management 2014; 79(1): 25-33.

[23] Şengül Ü, Eren M, Shiraz SE, Gezder V, Şengül AB. Fuzzy TOPSIS method for ranking renewable energy supply systems in Turkey. Renewable Energy 2015; 75(1): 617-625.

[24] Çelikbilek Y, Tüysüz F. An integrated grey based multi-criteria decision making approach for the evaluation of renewable energy sources. Energy 2016; 115(3): 1246-1258.

[25] Sağır H, Doğanalp B. Bulanık çok kriterli karar verme perspektifinden Türkiye için enerji kaynakları değerlendirmesi. Kastamonu Üniversitesi İktisadi ve İdari Bilimler Fakültesi Dergisi 2016; 11(1): 233-256.

[26] Balin A, Baraçli H. A fuzzy multi-criteria decision making methodology based upon the interval type-2 fuzzy sets for evaluating renewable energy alternatives in Turkey. Technological and Economic Development of Economy 2015; 23(5): 742-763.

[27] Büyüközkan G, Güleryüz S. Evaluation of Renewable Energy Resources in Turkey using an integrated MCDM approach with linguistic interval fuzzy preference relations. Energy 2017; 123: 149-163.

[28] Çolak M, Kaya İ. Prioritization of renewable energy alternatives by using an integrated fuzzy MCDM model: A real case application for Turkey. Renewable and Sustainable Energy Reviews 2017; 80: 840-853.

[29] Özcan EC, Ünlüsoy S, Eren T. ANP ve TOPSIS yöntemleriyle türkiye'de yenilenebilir enerji yatirim alternatiflerinin değerlendirilmesi. Selçuk Üniversitesi Mühendislik, Bilim ve Teknoloji Dergisi 2017; 5(2): 204-219.

[30] Özkale C, Celik C, Turkmen AC, Cakmaz ES. Decision analysis application intended for selection of a power plant running on renewable energy sources. Renewable and Sustainable Energy Reviews 2017; 70: 1011-1021.

[31] Boran FE. A new approach for evaluation of renewable energy resources: A case of Turkey. Energy Sources, Part B: Economics, Planning, and Policy 2018; 13(3): 196-204.

[32] Büyüközkan G, Karabulut Y, Güler M. Strategic Renewable Energy Source Selection for Turkey with Hesitant Fuzzy MCDM Method. In: Kahraman C, Kayakutlu G, editors. Energy Management—Collective and Computational Intelligence with Theory and Applications, vol. 149. Cham: Springer International Publishing; 2018. p. 229-250.

[33] Karaca C, Ulutaş A. Entropi ve Waspas yöntemleri kullanılarak Türkiye için uygun yenilenebilir enerji kaynağının seçimi. Ege Academic Review 2018; 18(3): 483-494.

[34] Engin O, Sarucan A, Baysal ME. Türkiye için çok kriterli karar verme yöntemleri ile yenilenebilir enerji alternatiflerinin analizi. International Journal of Social and Humanities Sciences Research (JSHSR) 2018; 5(23): 1223-1231.

[35] Toklu MC, Taşkın H. A fuzzy hybrid decision model for renewable energy sources selection. International Journal of Computational and Experimental Science and Engineering 2018; 4(1): 6-10.

[36] Karakaş E, Yıldıran OV. Evaluation of renewable energy alternatives for Turkey via modified fuzzy AHP. International Journal of Energy Economics and Policy 2019; 9(2): 31-39.

[37] Derse O, Yontar E. SWARA-TOPSIS yöntemi ile en uygun yenilenebilir enerji kaynağinin belirlenmesi. Endüstri Mühendisliği 2020; 31(3): 389-419.

[38] Yilan G, Kadirgan MN, Çiftçioğlu GA. Analysis of electricity generation options for sustainable energy decision making: The case of Turkey. Renewable Energy 2020; 146: 519-529.

[39] Solangi YA, Tan Q, Mirjat NH, Valasai GD, Khan MWA, Ikram M. Analyzing renewable energy sources of a developing country for sustainable development: An integrated fuzzy based-decision methodology. Processes 2020; 8(7): 825.

[40] Deveci K, Cin R, Kağızman A. A modified interval valued intuitionistic fuzzy CODAS method and its application to multi-criteria selection among renewable energy alternatives in Turkey. Applied Soft Computing 2020; 96: 106660.

[41] Karatop B, Taşkan B, Adar E, Kubat C. Decision analysis related to the renewable energy investments in Turkey based on a Fuzzy AHP-EDAS-Fuzzy FMEA approach. Computers & Industrial Engineering 2021; 151: 106958.

[42] Bilgili F, Zarali F, Ilgün MF, Dumrul C, Dumrul Y. The evaluation of renewable energy alternatives for sustainable development in Turkey using intuitionistic fuzzy-TOPSIS method. Renewable Energy 2022; 189: 1443-1458.

[43] San Cristóbal JR. Multi-criteria decision-making in the selection of a renewable energy project in spain: The Vikor method. Renewable Energy 2011; 36(2): 498-502.

[44] Yi SK, Sin HY, Heo E. Selecting sustainable renewable energy source for energy assistance to North Korea. Renewable and Sustainable Energy Reviews 2011; 15(1): 554-563.

[45] Sadeghi A, Larimian T, Molabashi A. Evaluation of renewable energy sources for generating electricity in province of Yazd: a fuzzy MCDM approach. Procedia-Social and Behavioral Sciences 2012; 62: 1095-1099.

[46] Mourmouris JC, Potolias C. A multi-criteria methodology for energy planning and developing renewable energy sources at a regional level: A case study Thassos, Greece. Energy Policy 2013; 52: 522-530.

[47] Stojanović M. Multi-criteria decision-making for selection of renewable energy systems. Safety Engineering 2013; 3(2): 115-120.

[48] Ahmad S, Tahar RM. Selection of renewable energy sources for sustainable development of electricity generation system using analytic hierarchy process: A case of Malaysia. Renewable Energy 2014; 63: 458-466.

[49] Troldborg M, Heslop S, Hough RL. Assessing the sustainability of renewable energy technologies using multi-criteria analysis: Suitability of approach for national-scale assessments and associated uncertainties. Renewable and Sustainable Energy Reviews 2014; 39: 1173-1184.

[50] Tasri A, Susilawati A. Selection among renewable energy alternatives based on a fuzzy analytic hierarchy process in Indonesia. Sustainable Energy Technologies and Assessments 2014; 7: 34-44.

[51] Al Garni H, Kassem A, Awasthi A, Komljenovic D, Al-Haddad K. A multicriteria decision making approach for evaluating renewable power generation sources in Saudi Arabia. Sustainable Energy Technologies and Assessments 2016; 16: 137-150.

[52] Afsordegan A, Sánchez M, Agell N, Zahedi S, Cremades LV. Decision making under uncertainty using a qualitative TOPSIS method for selecting sustainable energy alternatives. International Journal of Environmental Science and Technology 2016; 13: 1419-1432.

[53] Algarín CR, Llanos AP, Castro AO. An analytic hierarchy process based approach for evaluating renewable energy sources. International Journal of Energy Economics and Policy 2017; 7(4): 38-47.

[54] Ishfaq S, Ali S, Ali Y. Selection of optimum renewable energy source for energy sector in Pakistan by using MCDM approach. Process Integration and Optimization for Sustainability 2018; 2: 61-71.

[55] Yuan J, Li C, Li W, Liu D, Li X. Linguistic hesitant fuzzy multi-criterion decision-making for renewable energy: A case study in Jilin. Journal of Cleaner Production 2018; 172: 3201-3214.

[56] Rani P, Mishra AR, Pardasani KR, Mardani A, Liao H, Streimikiene D. A novel VIKOR approach based on entropy and divergence measures of Pythagorean fuzzy sets to evaluate renewable energy technologies in India. Journal of Cleaner Production 2019; 238: 117936.

[57] Solangi YA, Tan Q, Mirjat NH, Valasai GD, Khan MWA, Ikram M. An integrated Delphi-AHP and fuzzy TOPSIS approach toward ranking and selection of renewable energy resources in Pakistan. Processes 2019; 7(2): 118.

[58] Rani P, Mishra AR, Mardani A, Cavallaro F, Alrasheedi M, Alrashidi A. A novel approach to extended fuzzy TOPSIS based on new divergence measures for renewable energy sources selection. Journal of Cleaner Production 2020; 257: 120352.

[59] Niu D, Zhen H, Yu M, Wang K, Sun L, Xu X. Prioritization of renewable energy alternatives for China by using a hybrid FMCDM methodology with uncertain information. Sustainability 2020; 12(11): 4649.

[60] Li X, Zhu S, Yüksel S, Dinçer H, Ubay GG. Kano-based mapping of innovation strategies for renewable energy alternatives using hybrid interval type-2 fuzzy decision-making approach. Energy 2020; 211: 118679.

[61] Chen T, Wang Y, Wang J, Li L, Cheng PF. Multistage decision framework for the selection of renewable energy sources based on prospect theory and PROMETHEE. International Journal of Fuzzy Systems 2020; 22: 1535-1551.

[62] Wang Y, Xu L, Solangi YA. Strategic renewable energy resources selection for Pakistan: Based on SWOT-Fuzzy AHP approach. Sustainable Cities and Society 2020; 52: 101861.

[63] Mohammed HJ, Naiyf AT, Thaer AJ, Khbalah SK. Assessment of sustainable renewable energy technologies using analytic hierarchy process. In: IOP Conference Series: Earth and Environmental Science. IOP Publishing 2021; 779(1): 12-38.

[64] Abdul D, Wenqi J, Tanveer A. Prioritization of renewable energy source for electricity generation through AHP-VIKOR integrated methodology. Renewable Energy 2022; 184: 1018-1032.

[65] Assadi MR, Ataebi M, Sadat Ataebi E, Hasani A. Prioritization of renewable energy resources based on sustainable management approach using simultaneous evaluation of criteria and alternatives: A case study on Iran's electricity industry. Renewable Energy 2022; 181: 820-832.

[66] Goswami SS, Mohanty SK, Behera DK. Selection of a green renewable energy source in India with the help of MEREC integrated PIV MCDM tool. Materials Today: Proceedings 2022; 52: 1153-1160.

[67] Li P, Xu Z, Wei C, Bai Q, Liu J. A novel PROMETHEE method based on GRA-DEMATEL for PLTSs and its application in selecting renewable energies. Information Sciences 2022; 589: 142-161.

[68] Saaty TL. The analytic hierarchy process (AHP). The Journal of the Operational Research Society 1980; 41(11): 1073-1076.

[69] Saaty TL. Decision making with the analytic hierarchy process. IJSSCI 2008; 1(1): 83.

[70] Dhami I, Deng J, Strager M, Conley J. Suitability-sensitivity analysis of nature-based tourism using geographic information systems and analytic hierarchy process. Journal of Ecotourism 2017; 16(1): 41-68.

[71] Duke JM, Aull-Hyde RA. Identifying public preferences for land preservation using the analytic hierarchy process. Ecological Economics 2002; 42(1-2): 131-145.

[72] Saaty TL. Theory and applications of the analytic network process: decision making with benefits, opportunities, costs, and risks. RWS Publications 2005.

[73] Kuru A, Akın B. Entegre yönetim sistemlerinde çok kriterli karar verme tekniklerinin kullanimina yönelik yaklaşımlar ve uygulamaları. Öneri Dergisi 2012; 10(38): 129-144.

[74] Mastrocinque E, Ramírez FJ, Honrubia-Escribano A, Pham DT. An AHP-based multi-criteria model for sustainable supply chain development in the renewable energy sector. Expert Systems with Applications 2020; 150: 113321.

[75] Saaty TL, Tran LT. On the invalidity of fuzzifying numerical judgments in the Analytic Hierarchy Process. Mathematical and Computer Modelling 2007; 46(7-8): 962-975.

[76] Damgacı E, Boran K, Boran FE. Sezgisel bulanık TOPSIS yöntemi kullanarak Türkiye'nin yenilenebilir enerji kaynaklarının değerlendirilmesi. Politeknik Dergisi 2017; 20(3): 629-637.

[77] Karakul AK. Bulanık AHP yöntemi ile yenilenebilir enerji kaynağı seçimi. Bingöl Üniversitesi Sosyal Bilimler Enstitüsü Dergisi (BUSBED) 2020; 10(19): 127-150.

[78] Karaca C, Ulutaş A, Eşgünoğlu M. Türkiye'de optimal yenilenebilir enerji kaynağının COPRAS yöntemiyle tespiti ve yenilenebilir enerji yatırımlarının istihdam artırıcı etkisi. Maliye Dergisi 2017; 172: 111-132.

[79] Doğan H, Uludağ AS. Yenilenebilir enerji alternatiflerinin değerlendirilmesi ve uygun tesis yeri seçimi: Türkiye'de bir uygulama. Ekonomik ve Sosyal Araştırmalar Dergisi 2018; 14(2): 157-180.

[80] Karaaslan A, Aydın S. Yenilenebilir enerji kaynaklarının çok kriterli karar verme teknikleri ile değerlendirilmesi: Türkiye örneği. Atatürk Üniversitesi İktisadi ve İdari Bilimler Dergisi 2020; 34(4): 1351-1375.

[81] Albayrak ÖK. Yenilenebilir enerji kaynaklarının değerlendirilmesinde kullanılan çok kriterli karar verme teknikleri ve değerlendirme kriterlerinin incelenmesi: 2017-2020. Atatürk Üniversitesi İktisadi ve İdari Bilimler Dergisi 2020; 34(4): 1287-1310.

[82] Alkan Ö, Albayrak ÖK. Ranking of renewable energy sources for regions in Turkey by fuzzy entropy based fuzzy COPRAS and fuzzy MULTIMOORA. Renewable Energy 2020; 162: 712-726.

[83] Ulutaş BH. Determination of the appropriate energy policy for Turkey. Energy 2005; 30(7): 1146-1161.

[84] Kaya T, Kahraman C. Multicriteria decision making in energy planning using a modified fuzzy TOPSIS methodology. Expert Systems with Applications 2011; 38(6): 6577-6585.

[85] Erol Ö, Kılkış B. An energy source policy assessment using analytical hierarchy process. Energy Conversion and Management 2012; 63: 245-252.

[86] Yücenur GN, Çaylak Ş, Gönül G, Postalcıoğlu M. An integrated solution with SWARA&COPRAS methods in renewable energy production: City selection for biogas facility. Renewable Energy 2020; 145: 2587-2597.

[87] Kayakutlu G, Ercan S. Regional Energy Portfolio Construction: Case Studies in Turkey. In: Cucchiella F, Koh L, editors. Sustainable Future Energy Technology and Supply Chains. Cham: Springer International Publishing; 2015. p. 107-126.

[88] Topcu YI, Ulengin F. Energy for the future: An integrated decision aid for the case of Turkey. Energy 2004; 29(1): 137-154.